\documentclass[journal]{vgtc} 

\onlineid{0}

\vgtcpapertype{Application/Design Study}

\title{ DITTO: A Visual Digital Twin for Interventions and Temporal Treatment Outcomes in Head and Neck Cancer }

\author{%
  \authororcid{Andrew Wentzel}{0000-0002-2003-2750},
  \authororcid{Serageldin Attia}{0000-0002-0046-4337},
  \authororcid{Xinhua Zhang}{0000-0002-4084-6022},
  \authororcid{Guadalupe Canahuate}{0000-0001-5873-5454},
  \authororcid{Clifton David Fuller}{0000-0002-5264-3994},\\
  and \authororcid{G.Elisabeta Marai}{0000-0002-7212-9669}
}

\authorfooter{
  \item
  	Andrew Wentzel, Xinhua Zhang, and Liz Marai are with the University of Illinois Chicago
  	E-mail: \{awentze2 | gmarai\}@uic.edu
  \item
  	G. Canahuate is with the University of Iowa
  \item S. Attia and C.D. Fuller are with the MD Anderson Cancer Center
}
\abstract{%
  Digital twin models are of high interest to Head and Neck Cancer (HNC) oncologists, who have to navigate a series of complex treatment decisions that weigh the efficacy of tumor control against toxicity and mortality risks. Evaluating individual risk profiles necessitates a deeper understanding of the interplay between different factors such as patient health, spatial tumor location and spread, and risk of subsequent toxicities that can not be adequately captured through simple heuristics. To support clinicians in better understanding tradeoffs when deciding on treatment courses, we developed DITTO, a digital-twin and visual computing system that allows clinicians to analyze detailed risk profiles for each patient, and decide on a treatment plan. DITTO relies on a sequential Deep Reinforcement Learning digital twin (DT) to deliver personalized risk of both long-term and short-term disease outcome and toxicity risk for HNC patients. Based on a participatory collaborative design alongside oncologists, we also implement several visual explainability methods to promote clinical trust and encourage healthy skepticism when using our system. We evaluate the efficacy of DITTO through quantitative evaluation of performance and case studies with qualitative feedback. Finally, we discuss design lessons for developing clinical visual XAI applications for clinical end users.
}


\teaser{
  \centering
  \includegraphics[width=\textwidth, alt={A screenshot of the computer interface with 4 labeled sections. (A) On the left, an input panel with buttons and free form text for patient details, with diagrams for tumor location on the bottom. (B) In the center, a series of temporal plots for survival, locoregional control, and distant control probability over time, for 4 different cases. A toggle at the top indicates there is a secondary view. (C) On the top-right, two bars show the percentage of confidence each of the two models has in their treatment recommendation. (D) On the middle and bottom right, a waterfall plot shows how each feature contributes to the model descriptions}]{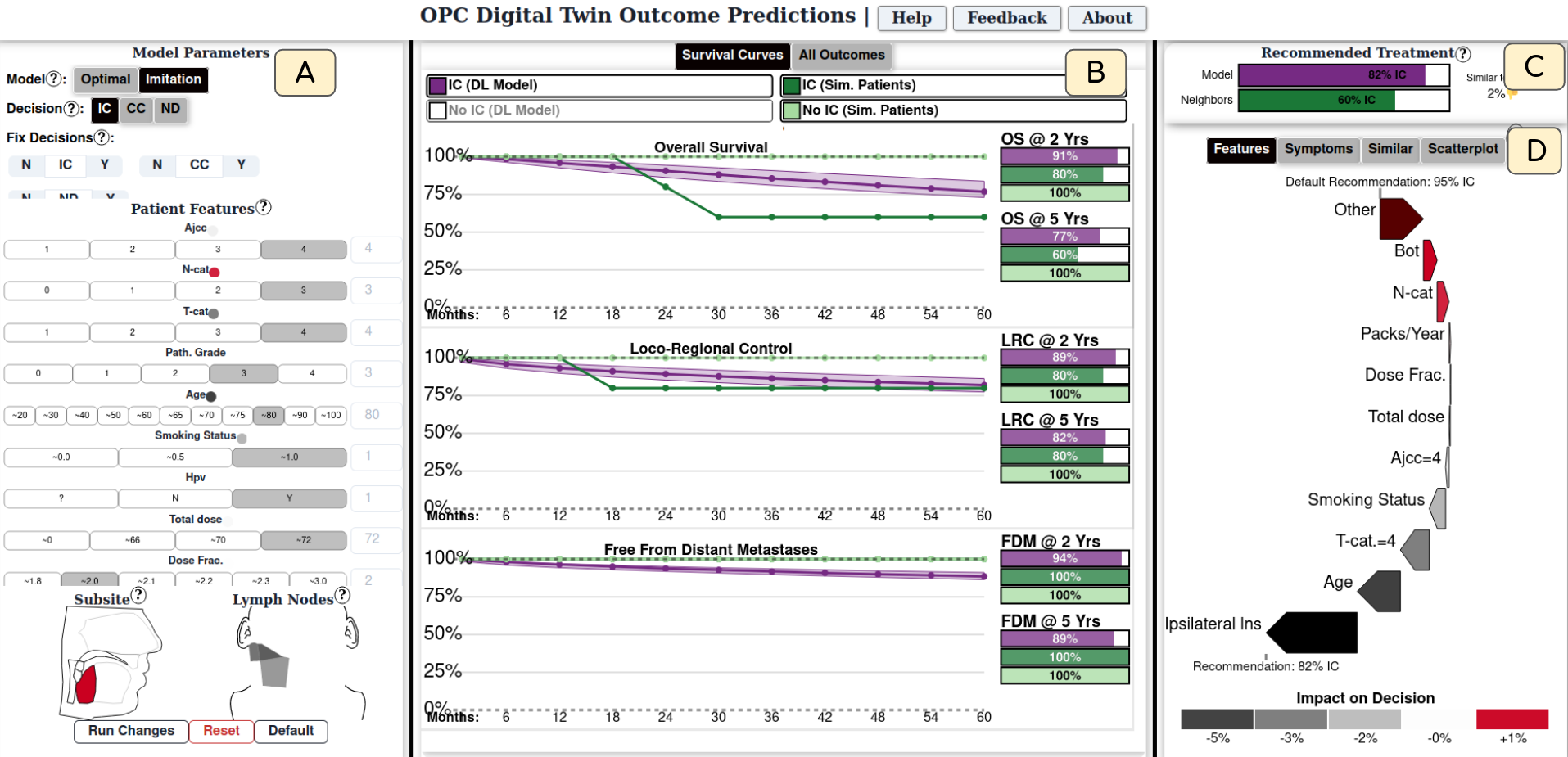}
  \caption{%
  	Overview of DITTO. (A) Input panel to alter model parameters and input patient features. (B) Temporal outcome risk plots for the patient based on different models and treatment groups. (C) Treatment recommendation based on the twin model and similar patients. (D) Auxiliary data panel, currently showing a waterfall plot of how each feature cumulatively contributes to the model decision.
  }
  \label{fig:dt_teaser}
}

\graphicspath{{figs/}{figures/}{pictures/}{images/}{./}} 

\usepackage{tabu}                      
\usepackage{booktabs}                  
\usepackage{lipsum}                    
\usepackage{mwe}                       

\usepackage{mathptmx}                  
\usepackage{multirow}
\usepackage{longtable}
\usepackage{url}

\begin{document}

\firstsection{Introduction}

\maketitle
Head and Neck Cancer (HNC) is a serious but treatable illness that affects up to 65,000 people each year in the United States alone. Care for HNC patients is a complex, multi-stage process that is dependent on the spatial location of the disease and its spread, and which includes potentially repeated cycles of surgery, chemotherapy, and radiation therapy. Determining the appropriate course of treatment for each patient is currently reliant on high level national guidelines and clinician cumulative experience. However, current guidelines do not adequately address the wide range of individual patient responses to treatments or the dynamic adjustments clinicians must make in response. For example, treating patients with chemotherapy before radiation treatment may reduce the overall tumor size and therefore reduce the risk of severe long-term side effects, but may also increase mortality risk. As a result, there is exceptional interest in digital twin (DT) models of the treatment process to help HNC oncologists better understand the potential risks and benefits of different treatment decisions at each state in the treatment process. Digital twins are data-driven simulations of patients and how they respond to treatment, which can be used to tailor treatments for individual patients based on how they are expected to respond to different interventions. DTs require complex simulations of a patient's health at multiple points in treatment, and thus rely on models that are more complex than those typically used in clinical settings (e.g., logistic regression). Data visualization is an underutilized resource that can help clinicians interact more effectively with these digital twins.

Visualization for digital twins for subject-matter experts is an under-explored visualization challenge~\cite{oral2024from}, with many additional challenges specific to HNC clinical decision-making. In terms of data, DTs consider multiple aspects of treatment, in addition to a combination of spatial and dynamic multivariate data to capture the patient state, which need to be visualized. In terms of outcomes, patient simulations yield dense, dynamic, and temporal outcome predictions, which need to be presented efficiently to users who may be interested in only a small subset of the resulting outcomes, depending on the context. 

Furthermore, creating usable DT models also constitutes a visual explainable AI (XAI) challenge. While many XAI approaches have been developed for explaining models to model builders, less work has looked at the specific needs of model clients, who have unique requirements when considering both model performance and model explanations. For example, HNC clinical decisions may heavily depend on factors like spatial features and clinician experience, making simplification of results difficult. Issues with model explainability and actionability may be a factor in the low penetration of ML models in medicine (<2\%~\cite{ahmad2018interpretable}) beyond medical image analysis ML. Additionally, since existing models often contain biased or insufficiently diverse datasets to perfectly model the cohort, it is important to give recommendations that allow for model introspection and support appropriate trust in the recommendations while allowing physicians to identify cases when the model should be disregarded. Finally, complex model results need to be communicated to physicians while ensuring that the visualizations are sufficiently familiar so that they require minimal training.

In this work, we introduce a visual analysis interface for digital twins in oropharyngeal cancer treatment (DITTO). Our specific contributions are: 1) Requirements engineering of the factors that HNC oncologists consider when interacting with digital twin systems for treatment planning; 2) The design and implementation of a visual computing system with a dual digital-twin back-end, one twin (set of models) of the HNC patients, and one twin (set of models) of the HNC physician decisions; 3) The design of visual encodings for the visual computing front-end, with a focus on supporting clinicians and supporting both trust and skepticism in the models; and 4) A qualitative evaluation of the system with clinicians, resulting in visual digital twin design insights.

\section{Related Work}
\subsection{Patient Risk Modeling}
Research in head and neck (HNC) oncology focuses on evaluating ways of improving patient outcomes through changes in treatment. Current approaches have seen relatively high survival rates ($\sim86\%$) in many HNC patients. As a result, current work often focuses on reducing side-effects (\textit{toxicities} or \textit{symptoms}) from treatment for patients with good survival probabilities. Earlier works have built interpretable models for predicting patient clinical outcomes for HNC patients such as survival and toxicity using clinical features~\cite{marai2018precision}, lymph node involvement~\cite{wentzel2021precision,luciani2020spatial}, tumor location~\cite{wentzel2019cohort,wentzel2020precision} and dose distributions~\cite{wentzel2023multi}, and radiomics~\cite{canahuate2023spatially}. This work is an extension of these approaches with a focus on temporally changing outcomes as well as intermediate treatment responses, which relies on more complex black-box models and post-hoc, instance based explanation methods for model interpretability.

Risk modeling for patients with censored time-to-event outcome data like survival~\cite{leung1997censoring} is generally modeled using approaches such as cox proportional hazard models~\cite{therneau1997extending}, non-parametric Kaplan-Meier analysis, and fully parametric models such as linear regression and survival trees~\cite{wang2019machine,zdilar2018evaluating}. This work adapts a deep-learning approach to survival modeling called deep survival machines (DSMs), which use a fully parametric mixture of distributions fitted to the training data~\cite{nagpal2021deep}. Other approaches have adapted deep learning approaches to Cox proportional hazard models~\cite{xu2023coxnam} and attention-based transformer models for predicting survival~\cite{li2022attention}. However, none of these models account for differences in patient response during treatment. 

In terms of Reinforcement learning, VA for interpretable RL is usually focused on targeting model builders~\cite{wang2019DQNViz,wang2022visual}. For clinical models, several systems have proposed attention weights for interpreting temporal neural networks~\cite{choi2016retain,ma2020concare}. In terms of visualization, RetainVis~\cite{kwon2019retainvis} focuses on exploring a recurrent neural network on temporal electronic health record data in patient cohorts. RMExplorer~\cite{kwonrmexplorer2022} uses subgroup statistics and feature attribution methods to explore model fairness in risk models.

More generally, DrugExplorer~\cite{wang2022extending} proposed a general framework for XAI applied to drug discovery. 
In terms of presenting models to users, Suh et al~\cite{suh2023are} and Zitek et al~\cite{zytek2022sibyl} discuss strategies for communicating models to domain experts, but do not expand this to applications in decision support. Kaur et al.~\cite{kaur2020interpreting} showed that many users can "over trust" erroneous model explanations they don't understand properly. VISPUR~\cite{teng2024vispur} discusses methods of identifying spurious correlations in causal models, but do not focus on integrating domain expert knowledge. 

\subsection{Digital Twins}
A digital twin is a digital model of a real-world system or process, that serves as the digital counterpart of it for practical purposes, such as simulation, integration, testing, monitoring, and maintenance. Although the term digital twin was introduced in 2010, visual steering of detailed computer simulations (i.e., fdigital twins) has been used before for flood simulation planning~\cite{waser2010world} and VR applications for manufacturing~\cite{zhu2019visualisation}. 

In healthcare, limited work has been done in exploring digital twins for patients using dashboards~\cite{lauer2023human,karabacak2023prognosis} and 3D models to visualize blood flow~\cite{mccullough2021towards}. Digital twin tools have also made for simulating physicians~\cite{talukder2022physicians}. Other approaches have built digital twins for radiation dosage adaptation~\cite{tseng2017deep}, glioblastoma treatment~\cite{zade2022deep}, and emergency department managementand~\cite{bouleux2023requirements}, but do not integrate visualization or explainability. Marai et al~\cite{G._Elisabeta_Marai_et_al_2019} developed a web visualization tool for HNC patient risk based on similar patients that allows for what-if analysis. Our work uniquely integrates visualization for both a digital twin and digital physicians. Additional, to our knowledge, there has been no interactive visual computing approach for digital twins that can also factor temporal decision-making.

\subsection{Decision Support Systems}
Relevant to this work is clinical decision support (CDSS) systems. Jacobs et al. discuss a CDSS for clinical depression~\cite{jacobs2021may}. Other systems have focused on identifying ways of supporting physician workflows for heart implants~\cite{yang2019unremarkable}, critical care patients~\cite{zhang2022get} and diabetes care~\cite{burgess2023healthcare}. Other work has focused on model building for CDSS Bayes networks~\cite{muller2023visual}, and integrating feature explanations to help train physicians in diagnostics~\cite{ouyang2024leveraging}. More generally, a recent study has suggested that users are more likely to use AI recommendations for harder tasks~\cite{ha2024guided}. Despite this, few visual systems have focused on decision recommendation in the context of explainable ML recommendations.  

Several systems have been developed specifically to visually communicate risk prediction to clinical users or patients, although none of them focus on deep learning-driven personalized patient outcomes. A majority of these systems focus on variants of Kaplan Meier plots to communicate patient survival based on general diagnostic features~\cite{corvo2019survivis,van2023head,chu2018survival}. Oncofunction~\cite{zebralla2020obtaining} focuses on helping patients plan post-treatment symptoms. PROACT~\cite{hakone2018proact} found patients were primarily interested in time left and survival risk at different time points using simple visualization methods. Vromans et al.~\cite{vromans2022need} found that some information seeking was a coping mechanism for a percentage of the population, and personal quality-of-life measures were equally as important as patient survival. Floricel et al~\cite{floricel2024roses,floricel2022thalis} uses temporal glyphs and Sankey diagrams to show clusters of patient symptoms over time. 
Other common tools have used Kaplan-Meier curves~\cite{van2023head} and nomograms~\cite{gafita2021nomograms} for HNC and prostate cancer. However, as far as we know, no online systems yet include digital twins with temporal state outcomes, or use model explainability methods with patient-specific predictions.

\section{Methods}

\begin{figure*}[ht]
    \centering
    \includegraphics[width=.9\textwidth, alt={Flowchart of the treatment sequence simulated by the D R L models. Intermediate results of induction and concurrent chemotherapy are used as inputs into the next decision. Final outcomes are a mixture of time-to-event curves and fixed binary outcomes. The policy model is trained to make the optimal decisions to based on the final outcomes.}]{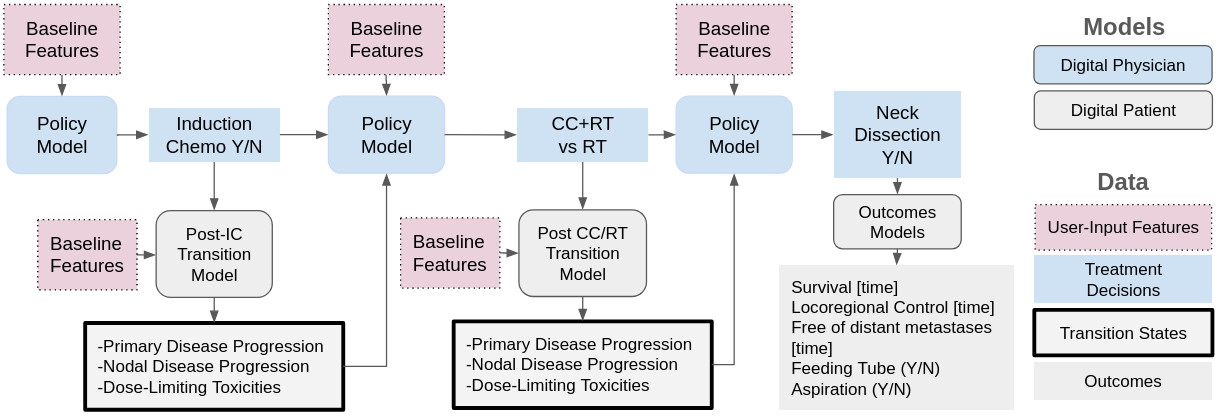}
    \caption{Overview of the treatment sequence simulated by the digital twin models, along with the features to be considered. Intermediate results of induction and concurrent chemotherapy are used as inputs into the next decision. Final outcomes are a mixture of time-to-event curves and fixed binary outcomes. The DT model is trained to make optimal decisions with respect to the final outcomes.}
    \label{fig:dt_treatmentsequence}
\end{figure*}


\begin{figure*}[ht]
    \centering
    \includegraphics[width=.9\textwidth, alt={Diagram of the process for filtering similar patients with similar likelihood to be treated, which are then used to estimate treatment outcomes. We use embeddings from the policy model to find a set of similar patients based on distance, and the further filter by predicted likelihood of receiving treatment for both treated and untreated similar patients. These simlar patients are then used to predict the likely outcomes for the patient with and without treatment.}]{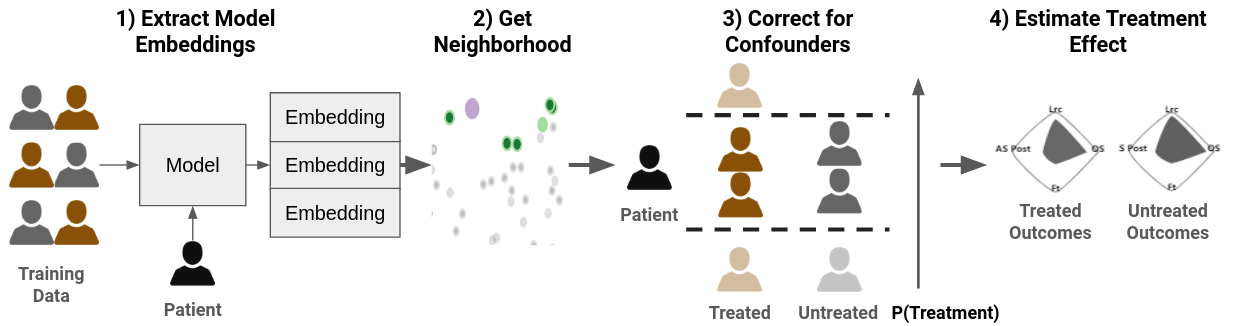}
    \caption{Diagram neighbor-based models when predicting patient outcomes. Model embeddings from the policy model are used to extract the most similar patients. Neighbors are filtered by their estimated likelihood of receiving treatment from the imitation model for those closest to the new patient. the difference between untreated and treated filtered neighbors can then be used to estimate impact of treatment.}
    \label{fig:dt_atediagram}
\end{figure*}

\begin{figure}[ht]
    \centering
    \includegraphics[width=.9\linewidth, alt={Image of survival curves for a patient based on different models. (A) Legend with toggle-able models and outcomes, currently showing only treatment groups. (B) Survival plot for a patient, showing prediction with concurrent chemotherapy and 95\% CI based on the DSM model (purple) and similar patients (green), along with fixed probabilities at 2 and 5 years. (C) Alternative outcomes view showing tables of predicted probabilities for additional outcomes.}]{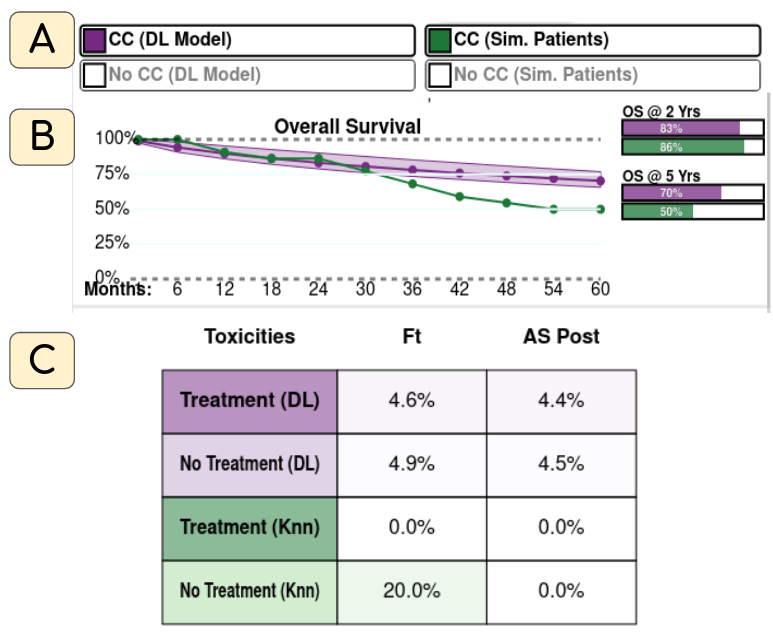}
    \caption{Image of survival curves for a patient based on different models. (A) Legend with toggle-able models and outcomes, currently showing only treatment groups. (B) Survival plot for a patient, showing prediction with concurrent chemotherapy and 95\% CI based on the DSM model (purple) and similar patients (green), along with fixed probabilities at 2 and 5 years. (C) Alternative outcomes view showing tables of predicted probabilities for additional toxicities (Ft -  Feeding Tube, AS Post - Aspiration Post-Treatment).}
    \label{fig:dt_outcomes}
\end{figure}

\begin{figure}[ht]
    \centering
    \includegraphics[width=.8\linewidth, alt={Truncated feature contribution waterfall plot showing how each feature contributes to the final model recommendation, relative to the default (median) patient. Color double-encodes attributions using a black for more negative values, and red for more positive values, and white for zero.}]{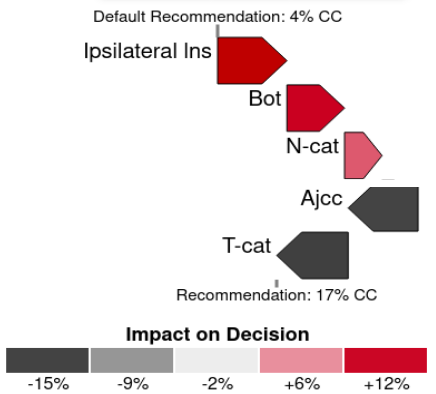}
    \caption{Truncated feature contribution waterfall plot showing how each feature contributes to the final model recommendation, relative to the default (median) patient. Color double-encodes attributions}
    \label{fig:dt_attributions}
\end{figure}

\begin{figure}[ht]
    \centering
    \includegraphics[width=.97\linewidth, alt={Example of a row in the similar patients view, showing the details of the average treatment group. 6 items are show: a diagram with areas corresponding to dose-limiting toxicites from chemo, a heat map of lymph node involvement based on an anatomical diagram of the neck, a heat map of tumor subsites based on an anatomical side view of the head, a radial chart of 4 staging variables, a radial chart of 6 demographic variables, and a radial chart with static patient outcomes: feeding tube, aspiration, 4 year survival, 4 year locoregional control, and 4 year distant control. Additional blue lines on the raidal charts indicate the input features of the current patient. All heat maps use a greyscale color scheme}]{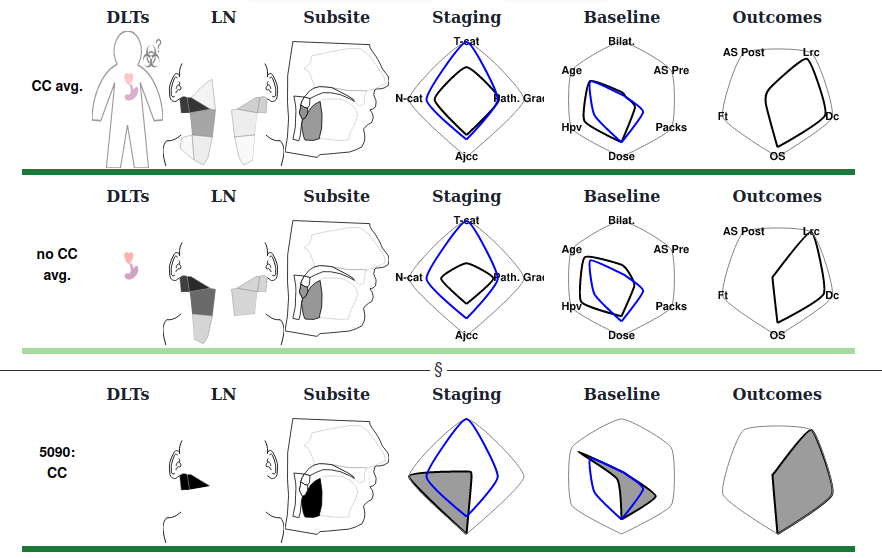}
    \caption{Similar patients view showing the row for average treatment group.  Each row shows toxicities, lymph node involvement, tumor subsite, staging, demographics, 4 year outcomes. Blue lines indicate the input features of the current patient in the staging and baseline Kiviat diagrams.}
    \label{fig:dt_similarpatients}
\end{figure}


\subsection{Requirement Analysis}
 This project was developed as part of a multi-year collaborative research project between HNC oncology radiotherapists from the MD Anderson Cancer Center, machine learning experts at the University of Illinois Chicago and University of Iowa, and visualization researchers at the University of Illinois Chicago.  We followed the Activity-Centered Design (ACD) methodology, which has higher success rates  in interdisciplinary settings than Human-Centered Design (63\% vs. 25\%) based on a survey of design studies~\cite{marai2017activity}. Requirements for both the interface and models were initially gathered through interviews with three research oncologists, and gradually refined during weekly meetings through multiple rounds of parallel prototyping and feedback over the course of several months. 

To gather formative feedback, a version of the interface designed based on our initial requirements was presented to a group of 11 physicians with clinical experience within the HNC oncology group at the MD Anderson Cancer Center. Several participants were familiar with the underlying dataset, but none had participated in the design of DITTO. During the session, participants were given an overview of the system components before being given a demo of the system and an online link where they were allowed to interact with the system on their own. This was followed by an open-ended discussion and a feedback interview.

\subsection{Data Abstraction}
Our dataset uses the patient cohort described in Tardini et al.~\cite{tardini2022optimal}. The cohort consists of 526 anonymized patients with squamous cell oropharyngeal tumors treated using definitive radiation therapy at the MD Anderson Cancer Center between 2003 and 2013. All data were collected after approval from the MDACC IRB (PA16-0303 and RCR03-0800). All patients included also had either recorded deaths or a minimum followup time of 4 years. Patients diagnostic data, treatment sequence, and outcomes were collected using EHR records. 

Standard treatments for patients include a mixture of \textit{surgery}, \textit{chemotherapy}, and \textit{radiation therapy} (RT). Chemotherapy can either be given before RT (induction - IC) or with RT (concurrent - CC). While real treatment can involve multiple rounds of each therapy, our simplified treatment sequence models the treatment process as 3 decisions: chemotherapy before RT (IC), chemotherapy concurrent with RT (CC), and neck-dissection (ND), a common surgery. These decisions are critical decision points, aligned with the standard-of-care~\cite{cdctreatment}. The entire treatment sequence model is shown in \cref{fig:dt_treatmentsequence}. Baseline features include age (cont.), if the patient is male or female/nonbinary (binary); race (binary x3); which regions of the neck have affected lymph nodes~{gregoire2014delineation} (binary x14); smoking status (never, former, current) (ord.); total radiation dose to the tumor (cont.) and dose per-visit (cont.); tumor staging (ord. x3); and tumor subsite (categorical x6). Tumor staging features (T, N, and AJCC) are ordinal rankings of tumor severity used to decide on treatment regime based on tumor size and spread for the main tumor and nearby lymph nodes~\cite{amin2016ajcc}. Race used a simplified grouping of demographics: White, African American/Black, Hispanic, and "Other", which is modeled as three one-hot variables in the data input. Minority inclusion reflects the demographics of the MD Anderson Cancer Center patient population which is approximately 85\% Caucasian and 15\% minority. Default gender is denoted as "male" in the model, to reflect the demographics of the patient population which is approximately 30\% females and 70\% males and did not have information on nonbinary individuals.  
 
 Each feature is associated with a feature-importance during modeling, based on how much it contributes to the twin model final decision, which is a value between 0 and 1. 

After each decision, the patient response to treatment is modeled as a transition state in terms of the response (change in size) of the primary tumor, nodal tumors, and any dose-limiting toxicities (DLTs). Tumor response is categorized into 4 groups based on the amount of change in tumor size: progressive disease, stable disease, partial response, and complete response. DLT types considered in our model are: Hematological, Neurological, Dermatological, Gastrointestinal, or Other. All DLT categories with fewer than 3 instances in the dataset are grouped into the "other" class.

For temporal outcomes, we consider patient survival (OS), local-regional control (LRC), and distant control (FDM). For each of these, we collected whether the event occurred, as well as either the time of the event or last follow-up date. Additionally, we recorded whether the patient was hospitalized for a feeding tube (FT) or lung aspiration (AS) within 6 months after finishing treatment as binary toxicity outcomes. As an auxiliary outcome, we extracted symptom ratings from a separate dataset of 937 patients with self-reported outcomes after receiving radiation therapy~\cite{wang2023improving}, which is used in a secondary view to display possible symptom trajectories.

\subsection{Digital Twins and Planning for Trust and Skepticism}\label{dt_details}

One of our goals is to provide support for both trust and skepticism in the system recommendations. In prior work~\cite{wentzel2023dass}, we have discussed visualizing "counterfactuals", where the model recommendation and ground truth diverge, and adding cues to highlight when model predictions should be given more scrutiny. Because DITTO aims to provide treatment recommendations for a new patient, where the ground truth is not available, we implement instead "neighborhood-based models" that are shown alongside the treatment recommendation and predicted outcomes, to encourage both trust and skepticism in the digital twin recommendations. These neighbor-based models show outcomes from similar patients in the cohort and are described in~\cref{ate}.

Our core dual digital twin system is based on modeling of patient responses at each time point for a given patient, alongside modeling of the physician-recommended treatment. We specifically refer to the patient response models as the "Patient Simulator", and the predicted physician treatment decisions as the "Policy Model". 

To further encourage trust and skepticism, and avoid reinforcing clinical bias, we planned to leverage and show recommendations from two deep learning models for the twin. In the context of modeling a physician we implement two approaches: imitation learning~\cite{zheng2022imitation}, which attempts to mimic what an expert would learn, and Deep Q Learning (DQN)~\cite{van2016deep}, which attempts to find an optimal decision based on expected future losses. e also implement a preliminary imitation learning model for use by clinicians. We refer to the DQN strategy as the "Optimal Policy Model" and to the imitation learning strategy as the "Imitation Policy Model". For the purpose of the interface, viewers can select the specific strategy to be used, and examine results from that model strategy. These two supervised deep learning models (DQN and imitation) are "Digital Twins" of the physician decision process, in addition to the patient simulator.

In total, DITTO leverages and can show three recommendations: two based on the DTs of the physician decision process, and one based on the neighborhood models.

\subsection{Task Analysis} \label{tasks}
Our system aims to help HNC oncology radiotherapists better understand the likely tradeoffs of adding other treatments to radiation therapy. Based on interviews, we found that clinicians generally consider treatment decisions at each stage individually, with a primary focus on identifying potential outcomes in terms of both immediate disease response, toxicity risk, and overall temporal outcomes up to 5 years. Individual interests, degrees of information seeking, and visual literacy varied based on the individual practitioner and their backgrounds. As a result, our system was designed to have flexibility, with the most prominent views being presented by default, and more detailed views available on demand.

Additionally, we have found during our collaboration that some clinicians trust their past clinical experience over neural networks and cohort-based reasoning, while others tend to trust the model even when the system does not make sense. As a result, our main design focuses on simultaneously showing results from both the supervised deep learning models used in the digital twin and neighbor-based models that use similar patients in the cohort (Sec.~\ref{ate}), to cue the user to have appropriate trust and skepticism in the system. 

Based on interviews and clinical feedback during the prototyping stage, we arrived at the following task abstraction:

\vspace{.2em}
\textbf{T1.} Identify the risk profile of a patient given a treatment selection.
\begin{enumerate}
\item \vspace{-0.1em} Display the temporal risk of negative outcomes for the individual patient using the digital twin
\item \vspace{-0.3em} Identify the cumulative patient risk in terms of the cohort of similar patients in the dataset
\item \vspace{-0.3em} Identify the ideal treatment plan for the patient
\item \vspace{-0.3em} Compare the patient to similar patients based on treatment and diagnostic data
\item \vspace{-0.3em} Display expected patient symptom profiles after radiation therapy
\end{enumerate}

\vspace{-0em}
\textbf{T2.} Identify relative benefit of treatment at the given time point.
\begin{enumerate}
\item \vspace{-0.1em} Display the potential gain in therapeutic efficacy in terms of survival, disease control, and additional side effects 
\item \vspace{-0.3em} Compare expected cumulative tumor control and survival to the probability of additional toxicity due to treatment for the patient 
\item \vspace{-0.3em} Display the risk of dose-limiting toxicity due to chemotherapy or treatment complications 
\end{enumerate}

\vspace{-0em}
\textbf{T3.} Identify the trustworthiness of the model predictions and recommended treatment
\begin{enumerate}
\item \vspace{-0.1em} Show the cumulative impact of each attribute on the recommended treatment in terms of percentage confidence
\item \vspace{-0.3em} Flag when the patient is an outlier in the cohort
\item \vspace{-0.3em} Display confidence intervals for the patient outcome predictions
\item \vspace{-0.3em} Compare the prediction of the DT and neighbor-based models
\end{enumerate}

\noindent\textbf{\emph{Nonfunctional Requirements}}
In addition to tasks, we determined a number of nonfunctional requirements. DITTO needed to build via visual scaffolding~\cite{marai2015visual} on encodings in existing clinical tools, such as Kaplan-Meier plots, barcharts, and cumulative distribution histograms.  
Additionally, several clinicians desired to be able to show these results to patients, and thus designs needed to avoid causing patent anxiety (i.e., scale survival should show risk always compared to 0). Finally, DITTO needed to be responsive and available online to be used by clinicians at any time, with minimal (< 5 seconds) time to produce results for a new patient.  

During the workshop, two participants requested information about data provenance and the model details, including limitations, available in the interface. Additionally, participants asked for the patient inputs to always be visible, and to only render additional views once an input has been manually submitted. Our original design also included both survival plots and barcharts of all outcomes and predicted transition states at the same time, in addition to median time to event for each temporal outcome. However, clinicians stated that most use-cases would focus on a smaller subset of results: the survival plots and survival at 2 and 5 years, with uncertainty values given, and that these designs should be centrally located, and additional results could be given on-demand as exact values.

\subsection{Deep Reinforcement Learning Models}
DITTO uses an extension of the dual digital twin system described in Tardini et al~\cite{tardini2022optimal}. The full system is shown in \cref{fig:dt_treatmentsequence}. 
Patients are assumed to follow a series of 3 binary decisions: Induction chemotherapy (IC), Concurrent chemotherapy (CC), and Neck Dissection (ND).  

Our digital twin is composed of multiple sub-models at each state in the treatment sequence, which are shown in more detail in \cref{fig:dt_treatmentsequence}. For the purpose of this section, we define terminology when referring to each of these sub-components. We call a model that predicts the patient's direct response to each treatment the "Transition Model", and the model that predicts long term temporal outcomes after definitive treatment is completed (i.e. survival and recurrence) the "Outcome Model". Following RL terminology, we refer to the model that simulates a physician as the "Policy Model". We have two versions of the policy model: The "Optimal Policy Model", and the "Imitation Policy Model", which attempts to predict the best treatment in terms of long term outcomes, and the treatment a physician would make, respectively. We only use one Policy model at a time, which is defined by the user. We use deep learning for all DT models due to their ability to deal with multimodal inputs with variable outputs and handle missing data~\cite{phung2019a}. 
The following section briefly discusses the details of each model.

To supplement the Digital Twin predictions, we show alternative predictions in the interface based on the most similar patients in the cohort at the given timepoint. We refer to this as the "Neighbor-based models" collectively, as we do not have to simulate responses at each step since all patients have ground truth decisions and patient responses available. 

Bellow we briefly describe each model. Due to space constraints, full details, model parameters, and evaluation can be found in the supplemental material Appendix A. 

\hfill \\
\noindent\textbf{\emph{Patient Simulator}}\\
To simulate the patient, we use a set of models to mimic intermediate response to treatment (transition models), and long-term response after treatment (outcome models). 

Transition models predict patient response to treatment in terms of tumor shrinkage and severe toxicities from treatment. Specifically, we consider primary disease response (PD), and nodal disease response (ND), which are each 4 categorical ordinal variables, as well as 5 binary results for different types of dose-limiting toxicities (DLTs). For  induction chemotherapy (IC), disease response is always assumed to be stable when no treatment is done.

For post-treatment outcomes, we predict a combination of temporal and static outcomes. We predict static outcomes using a deep neural network that predicts hospitalization due to two severe toxicities at up to 6 months after treatment: Aspiration (AS), and Feeding Tube insertion (FT). The temporal outcome model predicts cumulative patient risk over time for overall survival (OS), locoregional control (LRC), and distant metastases (FDM) for up to 5 years. Temporal risk models use a variant of deep survival machines (DSM)~\cite{nagpal2021deep}. For all three outcomes, the DSM model returns a mixture of parametric log-normal distributions for the patient that can be used to provide a cumulative survival risk over time.

Because clinicians listed confidence intervals as important for reasoning about the model predictions (T3.3), all transition and outcome models are trained using dropout on the penultimate layer between 50\% and 75\%. During evaluation, we re-run each prediction with random dropout at least 20 times, and then save the 95\% confidence intervals for each prediction.

\hfill \\
\noindent\textbf{\emph{Policy Modeling}}\\
The patient simulator models and ground truth responses are used as the environment to train a digital physician (policy model) 
. The policy model is a deep-learning based transformer encoder that predicts a binary treatment decision based on the baseline patient features, response to the previous treatment, previous decisions, and current timepoint. 

Because we need to explain the policy model recommendations (T3), we use integrated gradients~\cite{sundararajan2017axiomatic} to obtain feature importance for each decision relative to a baseline value. Integrated gradients was chosen as it satisfies the completeness axiom where attributions sum to the difference in the prediction between the baseline and actual recommendation, which was found to be easier to reason about with our clinicians. For our baseline, we assume the lowest possible rating for most ordinal attributes such as tumor staging or disease response, and the most common value for categorical attributes such as gender, ethnicity, and tumor subsite, as well as age and dose to the main tumor, based on feedback from clinicians and what they found most intuitive.

\subsection{Neighbor-based Models} \label{ate}
To provide an alternative model prediction to improve user trust (\cref{tasks}). we provide methods for estimating different patient outcomes using similar patients in the cohort, based on the embeddings taken from the final layer in the policy model for the given time-point and output. Our approach uses a modified variant of average treatment effect, which is used in causal modeling for finding predicted effects from treatment while correcting for confounders. 

For a new patient, we calculate a set of $k$ patients whose embeddings are most similar at each time point in terms of embedding using euclidean distance. When predicting treatment policy (physician choices), we use a smaller subset of the $n, n < k$ most similar patients and report the percent of patients that received treatment. For other outcomes and patient response, we take from the $k$ patients those with a predicted probability of receiving treatment that are within a certain value of the patient. We then calculate the relative prevalence of each outcome for the untreated and treated patients within this propensity-matched~\cite{austin2011introduction} group (\cref{fig:dt_atediagram}). For our system, 
We calculate the value difference as a fixed percentage of the standard deviation of the logits of the propensity scores in the cohort, defined as: 
$$cd = \alpha*\sqrt{\frac{1}{|X|} \sum_{x \in X} \bigg(\ln\big(\frac{p_{x}}{p_{x}-1}\big) - \frac{1}{|X|}\sum_{k \in X}\Big(\ln\big(\frac{p_{k}}{p_{k}-1}\big)\Big)\bigg)^{2}}$$
Where $X$ is the cohort and $p_{n}$ is the predicted probability of patient n receiving treatment. We use an $\alpha$ of .1 based on the suggested formula in~\cite{austin2011optimal}, which is increase in increments of .1 until treated and untreated groups have at least 5 patients. 

\subsection{Implementation}
Our back-end was implemented in Python using flask and pandas for data processing. Deep learning (digital twin) models use Pytorch, and deep survival machines use modified code taken from the auton-survival package~\cite{nagpal2022auton}. Feature attributions were calculated using the Captum package~\cite{kokhlikyan2020captum}. Our system front-end uses react with d3.js. Our online interface requires approximately 3.6-4.5 seconds to return simulation results for a new patient with two cores on an AMD EPYC 7452 Processor and requires 4GB of ram with 4 worker processes on the server, based on test queries for 10 random patients in the cohort. Specific model parameters where chosen via model tuning are given in the supplemental material.

\section{Design}
\subsection{Layout and Workflow}
Our main system is divided into three main components: input, patient outcomes, and treatment recommendation + supplemental views. First, an input panel on the left is used to change model and patient details (\cref{fig:dt_teaser}-A). To minimize cognitive load, we focus on only showing one, user-selected treatment (IC, CC, or ND) at a time. Users can optionally decide on other treatment decision when calculating future patient outcomes, with the policy model handling the other treatment decisions when nothing is input by the user.
Next, central views show patient survival outcomes (\cref{fig:dt_teaser}-B) as well as the recommended treatment for the patient (\cref{fig:dt_teaser}-C). Finally, additional views are shown via tabs to users who have an interest in more detailed information, such as model feature explanations (\cref{fig:dt_teaser}-D), similar patients, additional outcomes, and predicted symptoms ratings. These views are changed by toggling a set of buttons above the panel. Because many views are only of interest to certain users, we added functionality to resize width of each view via dragging the black vertical dividers, to allow users to expand auxiliary views as needed, while keeping the main goal of evaluating patient outcomes the main focus.

Whenever model predictions are shown in the interface, we present the deep-learning based Digital Twin predictions, and the neighbor based models. We use purple to encode Digital Twin predictions, and green to encode similar patient predictions. 

\subsection{User Input}\label{input}
The left panel allows inputting the relevant patient features and model parameters into the system. At the top, prompts are given for model input parameters: 1) whether the policy model should use the "optimal" or "imitation" strategy (\cref{dt_details}); 2) what decision is being considered; and 3) if any of the other decisions in the system are assumed to be "fixed" (yes or no). By default, the decision is decided by the currently selected policy model's recommendation.

Below the model parameter input is a panel for the current patient (\cref{fig:dt_teaser}-A).
By default, the average values for each feature are selected. We found that clinicians tend to think of continuous variables such as smoking pack-years and age in terms of discrete "bins" therefore, all features are shown using categorical stylized radio buttons to make selection easier, with free-text inputs on the side that allow users to use specific values when desired. These values are checked for validity based on the feature. When analyzing concurrent chemotherapy or neck-dissection, users can either specify the patient's primary and nodal tumor response to the previous round of chemotherapy, or allow the system to estimate this response automatically.

For the spatial inputs: affected lymph nodes and tumor subsites, we allow users to directly interact with diagrams of the respective areas. The diagram for the lymph nodes was previously developed alongside clinicians in our work with explainable lymph node clustering~\cite{wentzel2020explainable}. The diagram of tumor subsites were adapted from diagrams created by the MD Anderson Cancer Center. See the Appendix B for a labeled description of each spatial diagram.

In addition to feature input, we include color cues in the feature attribution plot for each of the features, described in detail in~\ref{recommendation} (T3.1). These are shown as colored dots next to each feature for nonspatial inputs, and as a color fill in the spatial features.

Because we do not want to re-run the computationally expensive simulation every time a feature or parameter is changed, a new simulation is run using the updated features once the user selects the "run changes'' button at the bottom. Additional buttons reset the feature inputs to the last time the simulation was run, and load the default patient features.

\subsection{Survival Plots and Outcomes}
When collecting feedback from HNC clinicians at the MD Anderson cancer center, several clinicians  suggested that users with less information seeking behavior will primarily be interested in seeing tumor control and survival risk for treated and untreated groups over time. As a result, we centrally place an outcomes view panel (\cref{fig:dt_outcomes}) that shows the model predictions for all relevant endpoints in our system. By default, we show temporal plots for survival, local-regional control, and distant metastasis for the treated and untreated groups using the Digital Twin outcome models (T1.1) and neighbor predictions (T1.2), up to 60 months post-treatment (\cref{fig:dt_outcomes}-B). We also include 90\% confidence intervals for Digital twin predictions as semi-transparent envelopes (T3.3). We chose to use temporal outcome plots as the main outcome plot, as oncologists often use variants of Kaplan Meier survival plots to assess patient risk. Additionally, the legend at the top can also be used to toggle off the visibility of certain models or treatment groups when the user only wants to see predictions for certain parameters (T3.4) (\cref{fig:dt_outcomes}-A). Each output is color-coded, where hue encodes model group (Digital twin vs neighbor-based) and luminance encodes treatment group (darker for treated groups).

Because a subset of information-seeking clinicians were interested in more details regarding patient response, an alternative window (\cref{fig:dt_outcomes}-C), shows static risk tables for all transition outcomes and temporal risk at 2 and 5 years for both Digital twin and neighbor-based predictions, for both the treated and untreated groups, for a total of 4 predictions each, via a toggle button (T2.1, T2.2, T2.3). This view relies on direct encoding of features. Additionally, each cell is color coded, with opacity encoding the risk percentage.  These additional results were originally encoded as a barchart shown alongside the survival plots, but were moved to a simpler, more explicit table shown on demand based on clinician feedback as well as recent findings suggesting that tables with explicit values are less prone to confirmation bias when reasoning about the data~\cite{xiong2022reasoning}.

\subsection{Treatment Recommendation} \label{recommendation}
The right panel of DITTO is devoted to more detailed model results, based on the varying requirements cited by different clinicians. We show the recommended treatment based on both the policy model, and similar patients at the top, in terms of a percentage between 0 and 100\% for the suggested treatment (\cref{fig:dt_teaser}-C) (T1.3). 
To provide a cue as to how reliable the model recommendation is, we calculate the Mahalanobis distance between the patient embedding taken from the model for each time point and the rest of the cohort (T3.2). We then calculate the relative percentile of the distance for this patient relative to the rest of the cohort (e.g., 0 to 100\%), which is shown next to the recommendation. We show a symbol (thumbs-up vs thumbs-down) based on if the percentile is below or above 75\%, respectively. This feature was based on a specific clinician request for a cue regarding whether the new patient recommendation can be trusted based on the cohort being used. Our original design included a full histogram. However, during the workshop, several clinicians misread the histogram, as some assumed being in the middle was better and others assumed the left was better. Additionally, clinicians did not find seeing the distribution of the full training cohort useful, and thus recommended using a text rating. 


Below the model recommendation, a panel shows additional custom model details. By default, the view shows a waterfall chart variant(\cref{fig:dt_attributions}). This view shows the cumulative impact of each attribute on the final decision in terms of percentage confidence in the given treatment on the x-axis (T3.1). The baseline shows the decision impact for a "default" patient, which is either the lowest possible value for ordinal (e.g., tumor staging) or continuous values, or the most common value for categorical features. We then show the impact of each feature as an error moving the decision along the x-axis. Because the integrated-gradients feature attribution method satisfies "completeness", the final position at the bottom is equal to the position of the final decision relative to the first decision point. Each bar is drawn as an error that uses a diverging color scheme to double-encode impact size. All values below a certain threshold (1\%) are aggregated into an "other" value as they have negligible interest to users. Features are shown in order of positive impact from the top to the bottom. This view was finalized as waterfall charts are an established method of showing feature attributions~\cite{gamut}, with the arrows and color encoding added to improve intuitiveness of the system. Additionally, it was very well received by clinicians during prototyping, and described as "very intuitive" by a collaborator with no prior experience with feature attributions.

\subsection{Similar Patients}
Based on interviews and previous experience with clinicians, many HNC oncologists are interested in using previous patients to reason about likely outcomes and the trustworthiness of the prediction and improve domain sense. As a result, we include an optional view that shows details on the similar patients used in the Average Treatment Effect estimates (\cref{fig:dt_similarpatients}) (T1.2, T1.4). The view shows feature summaries of each patient, as well as the average values for the treated and untreated groups. Each patient is encoded as a single row of patients. We show the tumor subsite and lymph nodes as heatmaps using the diagrams described in \cref{input}, as well as a diagram for and dose-limiting toxicity from the current treatment (see supplemental material). Additionally, we show three Kiviat charts with distributions of the most relevant features: diagnostic tumor staging (T-stage, N-stage, Overall Stage, and pathological grade), important clinical features (HPV, smoking status, age, etc.), and patient outcomes at 4 years (survival, local-regional control, distant control, aspiration, and feeding tube). The features for the current patient for non-outcome features are overlaid on top of each patient in blue, to support comparison between the groups and the current patient. This design was based on prior work showing promising results for diagram based spatial encodings~\cite{wentzel2023dass,wentzel2020explainable}, and radial charts to encode clinical features~\cite{marai2018precision,mantovani2023kiviat} when displaying similar patients for clinicians, along with positive feedback from collaborators.

We use colored borders and labels to indicate which patients are in the treated and untreated groups. This view is included as it was found to be useful for clinicians that value inspecting individual patients, or identifying confounders that may impact the recommendation of the neighbor-based predictions. However, since many clinicians said this functionality was only a secondary concern, it is hidden by default.

\subsection{Symptoms}

\begin{figure}[ht]
    \centering
    \includegraphics[width=\linewidth, alt={Plots of predicted patient symptom trajectories. Dark green indicates average of patients that receive a selected treatment, light green is average of patients that don't receive treatment. Faint lines indicate trajectories of the cohort patients used to make the prediction.}]{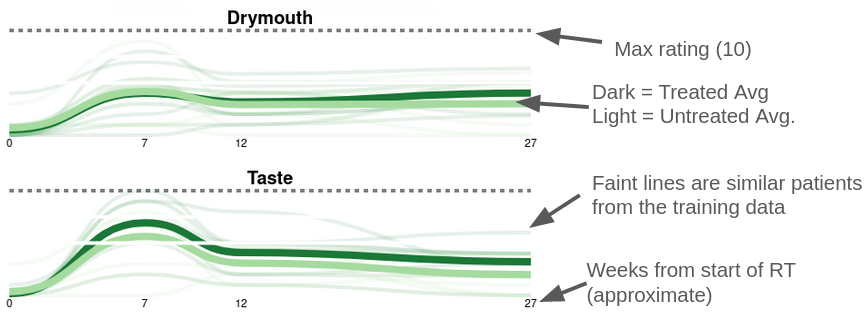}
    \caption{Symptoms prediction for a patient. Dark green indicates average of patients that receive a selected treatment, light green is average of patients that don't receive treatment. Faint lines indicate trajectories of the cohort patients used to make the prediction.}
    \label{fig:dt_symptoms}
\end{figure}

Finally, because several oncologists expressed a desire to see the effect of treatment on long-term subclinical side effects, we include a KNN-based symptom progression model for the patient (\cref{fig:dt_symptoms}) (T1.5). Due to data constraints, this view only includes a neighbor-based model, with no deep-learning based model as the cohorts were different and we were unable to get a sufficiently accurate model. This view shows self reported symptom progression for 10 different symptoms for a period of 6 months after the start of radiation treatment. Each similar patient is shown as a faint line, and group median values for treated and untreated groups are shown as bold lines. Symptoms are ordered by mean rating at the end of the time-period, as clinicians are most interested in long-term side effects that are more likely to be permanent.

\section{Qualitative Evaluation}
A quantitative evaluation of the models used in the system is included in the supplementary materials. To further evaluate DITTO, we performed two case studies with two users: one HNC clinician with 9 years of experience, 4 years of which were at the MD Anderson Cancer center, along with one Data Mining researcher, to find out how oncologists interact with the system. The case studies covered the evaluation of a single patient each and were performed via Zoom meetings with desktop sharing. To assess how different model recommendations might affect the users, we selected one patient that had both the neighbor-based and DT model agree with the true patient recommendation (non-counterfactual) and a case where the neighbor-based and DT disagreed with each other (counterfactual). The policy model strategy was set to "Imitation" based on clinician preference. Qualitative feedback was collected via a debriefing interview derived from the System Usability Scale~\cite{brooke1996sus} structure.

\subsection{Typical Recommendation}
\begin{figure*}[ht]
    \centering
    \includegraphics[width=.9\textwidth, alt={screenshots of interface during the first case study. (A) Feature Attributions and recommendation (truncated) showing that LN spread and Race are the main predictors of the patient receiving CC. (B) Patient survival curves. Green lines show very low survival for both treated (dark green) and untreated (light green) groups, but high survival from the DSM model (purple). (C) Survival curves for the patient when their race is changed to "white/caucasian". Similar patients have much higher survival rates.}]{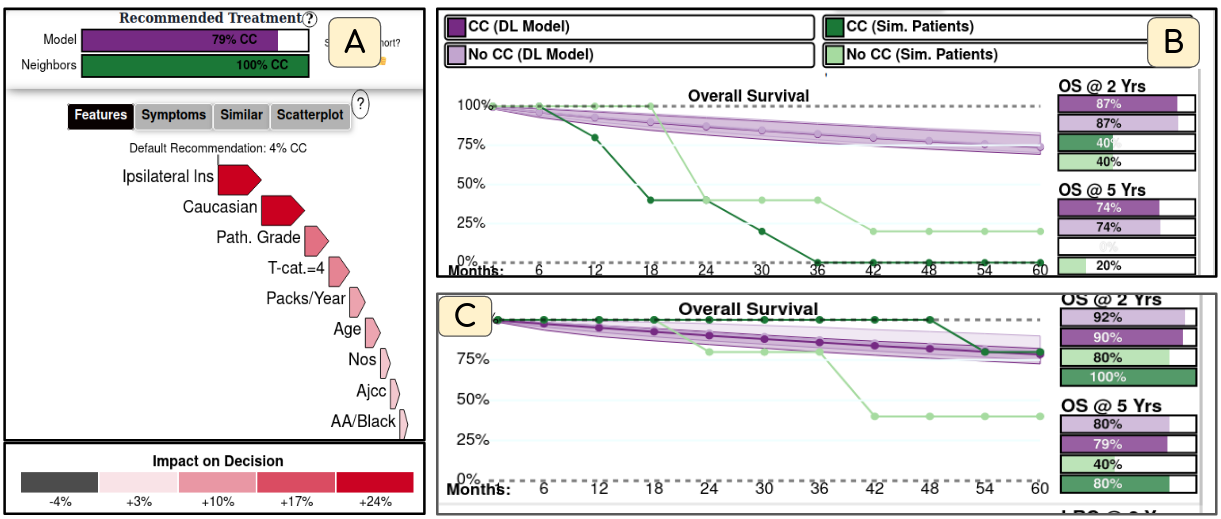}
    \caption{First case study. (A) Feature importance and recommendation (truncated) showing that LN spread and Race are the main predictors of the patient receiving CC. (B) Patient survival curves. Green lines show very low survival for both treated (dark green) and untreated (light green) groups, but high survival from the DSM model (purple). (C) Survival curves for the patient when their race is changed to "white/caucasian". Similar patients have much higher survival rates.}
    \label{fig:dt_casestudy1}
\end{figure*}

Our first case study was taken from an example patient where both the neighbor-based and Imitation policy model agreed with the clinical ground truth.
Starting with the patient input, the patient was notable for having a high T-stage (large primary tumor) and "Not Otherwise Specified" tumor location, suggesting that the patient had a large, irregularly positioned tumor, and being African American. The clinician then moved to the treatment recommendation (\cref{fig:dt_casestudy1}-A) to confirm that the model recommendation lined up with the similar patients in the cohort, where 100\% of patients receive chemotherapy. Looking at the feature importances for the policy model recommendation (T3.1), they noted that the most prominent features are the LN spread, the patient's race, the pathological grade, and the T-staging. While this finding mostly lined up with clinical reasoning, the impact of race was surprisingly high (+33\% chance of CC). 

In the survival outcomes (\cref{fig:dt_casestudy1}-B), they noted that there was a large discrepancy between the predicted survival, and those reported by the cohort (T3.4):  only 40\% of similar patients survived 2 years (T1.2), and none of the treated group survived 5 years despite a predicted survival rating of 89\% with high confidence (T1.1, T3.3). Interestingly, outcomes were better in the untreated group. Looking at the similar patients, we could see higher T-stage and pathological grade in the CC group, which may account for the difference, although it was unclear if race also impacted this (T1.4).

Looking back at the issue of race, the group tested this patient by changing only their race to "white/Caucasian". Indeed, this changes both the predicted treatment from the Deep Policy model (73\% no cc), and the patient outcomes, with significantly higher rates of predicted survival in the similar patients (\cref{fig:dt_casestudy1}-C) (T3.2). This led to a discussion on the use of race in the model, where we discussed issues of bias and confirmed that, indeed, race has an impact on physician treatment and patient outcomes, which requires further study~\cite{naghavi2016treatment}. Interestingly, the clinician also tested the "optimal" policy model, which only showed a minimal impact of race (< 1\%) on the recommended treatment, which was no CC. Notably, the optimal policy model listed the low pathological grade, tumor location, and AJCC stage as reasons to not give CC, while the LN spread is given as the primary reason to give CC. Based on the predicted outcomes, we noticed a much higher risk of side-effects (5.9\% chance increase in feeding tube and 3.6\% chance increase in Aspiration), with non-significantly higher predicted chance of tumor response or control, which led to the no-CC recommendation (T2.2, T2.3), as well as slightly higher incidence of severe symptoms in the symptom plot for the CC group (T1.5).

\subsection{Counterfactual Recommendation}
\begin{figure*}[ht]
    \centering
    \includegraphics[width=.9\textwidth, alt={screenshots of interface during the second case study. (A) Average of treated and untreated groups. Treated patients have lower LN spread and staging, and higher survival rates, which is counter-intuitive. (B) Feature attributions for ipsilateral LN levels II (top) vs levels II-IV (bottom), where dark red encodes higher likelihood of receiving CC. Changing the patient to have LN level IV involvement significantly increases confidence that the patient should receive CC.}]{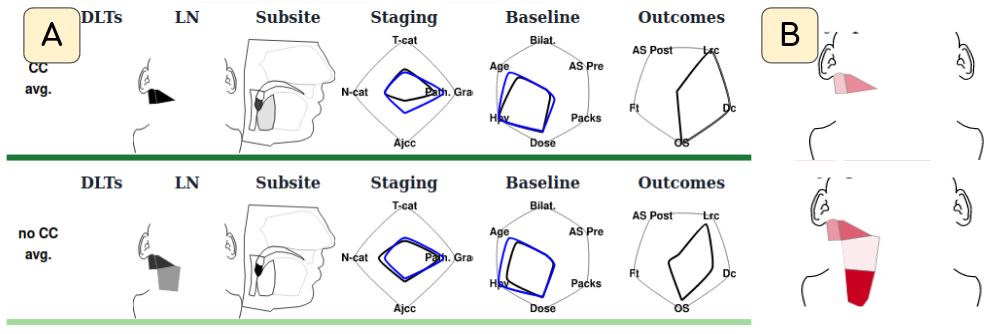}
    \caption{Second case study. (A) Average of treated and untreated groups. Treated patients have lower LN spread and staging, and higher survival rates, which is counter-intuitive. (B) Feature attributions for ipsilateral LN levels II (top) vs levels II-IV (bottom), where dark red encodes higher likelihood of receiving CC. Changing the patient to have LN level IV involvement significantly increases confidence that the patient should receive CC.}
    \label{fig:dt_casestudy2}
\end{figure*}

In this second case study, we examined a patient where the Deep Policy Model predicted no CC, while the most similar patients all received CC. In this case, the group noticed that patients had relatively low staging and low smoking, which the clinician confirmed lined up with the patient not needing CC in most cases (T1.3). They speculated the difference may be due to physician preference or other factors, such as features not accounted into the model but present in the lab notes (e.g., the patient having only one kidney). Additionally, they noted that in this case, the existing guidelines cite smoking and Lymph node levels as the main causal factors. We can see in the input LN diagram that the patient had LN levels II effects (\cref{fig:dt_casestudy2}-B, left), the most common levels, which have an impact on an increased chance of CC.

Noting in the outcome panel that there is a relatively high chance of survival for both groups given the low risk (T1.1, T1.2), and similar risk profile for treated and untreated groups (T2.2), the clinician moved to the similar patient panel. Notably, both treated and untreated groups had similar characteristics, but the untreated group actually had more nodal extension to level 3, higher staging, a higher average smoking rate, and lower survival and tumor control after 4 years (\cref{fig:dt_casestudy2}-A) (T1.4). They noted that this may confirm that the difference in the cohort treatment may be due to the physician or other factors.

Moving to the input panel, the clinician tested the impact of the two changes given by the physician: lymph node extension to level IV, and smoking > 20 pack-years, and confirmed that with these changes the model indeed changed to predict CC (52\% chance), and that LN level IV was a major factor in the change in treatment (\cref{fig:dt_casestudy2}-B) (T3.1).

\subsection{Qualitative Feedback}
Feedback from HNC clinicians at the MD Anderson Cancer Center was very positive, stating that the system was "really attractive'' and "amazing". When asked about their favorite features of the interface, multiple participants stated that they liked the views of similar patients, as well as symptom progression in the auxiliary panels. They also felt that many clinicians would be more interested in just the outcomes in the center. In response to this feedback, we turned the additional panels into a separate on-demand view. The participants also found the feature attributions interesting, saying "I also like the neighborhood panel and the multiple outcomes together". Some were particularly interested in the lymph node involvement levels for similar patients, as well as how this relates to feature importance in the policy model. They were also able to identify possible sources of data bias in the predictions by looking at the treated and untreated groups. When asked about the usefulness of the simplified three-decision model, the most senior clinician commented that "The 3 decision points are critical decision points, aligned with the standard of care. We could get more granular, but it's a great start."

\section{Discussion}
Our results show that DITTO is an effective tool for treatment planning for HNC clinicians using a novel Digital Twin system. Clinician feedback was very positive, with a variety of "favorite" components and background, suggesting that DITTO can handle a variety patient treatment goals. Additionally, while we had initial concern that the use of two model outputs would prove confusing, our case studies show that investigating model discrepancies indeed leads to interesting discussion into how patients should be treated and how physician preference or uncounted variables may impact certain recommendations. 

\subsection{Design Lessons}
A majority of explainable ML work has focused on visualizations meant for model builders or clinical researchers to use in research context, while most clinician-facing systems focus on relatively simple models~\cite{louise2023a,van2023head,chu2018survival}. In this regard, this system is a novel attempt to deliver the results of a complex Digital Twin system to clinical end users. In particular, this work focused on two challenges: delivering many potential results to clinicians in a way that allows them in a way that is relatively accessible, and to find a way to balance encouraging oncologists to use the system while not overly relying on potentially incorrect predictions. We list here the specific design insights we've developed during this participatory design process.

\emph{L1. Use visual scaffolding.} Our users were clinicians who had experience with risk modeling visualization. We found our best results by scaffolding, such as relying on temporal plots and spatial anatomical diagrams. Previous attempts at novel encodings such as histograms or unique glyphs were less successful with wider audiences.

\emph{L2. Account for different information-seeking needs.} We found in our interviews and literature reviews that the degree of information seeking behavior, as well as attitudes towards different models, varied greatly between clinicians. For example, we found that some users were completely uninterested in seeing similar patients or a scatterplot of the training cohort, while others listed the similar patients as their "favorite" part of the system and were able to identify potential confounder bias by looking at the similar patients. We also found that many users were primarily interested in seeing only the recommended treatment and time-to-survival, so this information could be communicated to patients, and felt additional features were distracting in the interface. As a result, we altered our design to afford these additional features in secondary tabs, and allowed for resizing of the different parts of the interface, while highlighting only the survival plots by default.

\emph{L3. Provide access to multiple models, and cues such as counterfactuals and confidence intervals to balance user expectations of the model.} In using our system, we found that clinicians have a tendency to either fully trust or distrust a model in the absence of additional cues, and expressed a desire for "honesty" in terms of model confidence. To encourage users to "think slowly"~\cite{kahneman2011thinking,kaur2020interpreting} about the model predictions, we relied on multiple cues: showing different model predictions side-by-side, using model confidence intervals when available, and placing the feature attribution plot prominently in the visualization. Still, there is necessarily a design tradeoff between interface simplicity, user acceptance of the model, and the number of additional cues. While many lay-users may prefer only being given a single prediction, we consider this a questionable design tradeoff with respect to XAI. As a result, we initially show users all model predictions, with the option to toggle off information.

\subsection{Limitations and Future Work}
In terms of limitations, our current models are limited by data availability and model performance. Our dataset requires modeling 19 different outcomes and transition state variables while relying on less than 600 patients from a single institution with limited demographic diversity. Furthermore, a more granular digital twin system could consider multiple rounds and dosages of chemotherapy and surgery. Our available dataset is also specific to oropharyngeal HNC patients. In terms of our interface, we focus on a limited group of HNC clinicians, a few of whom may have above-average visual literacy and information seeking behavior.

Our imitation learning model inherits existing treatment biases, and diversity shortcomings in the training data. While the initial goal of the system is to reveal these biases as shown in our case studies, there is the potential for users to interpret these explanations as justification for biased reasoning when over-trusting the system. Regardless, this bias should not be reflected in the risk prediction or optimal model, which should theoretically contradict the treatment recommendation in such a case. 

In terms of generalizability, the general approach can be applied to any similar treatment sequence that can be simplified into discrete decision stages, and a majority of our visualization system is domain-agnostic, except for the tumor subsite and lymph node spread diagrams. Regarding visualizations, our algorithms for uncertainty, feature attribution, and similarity are specific to deep learning classification, but these values can be obtained more generally through bootstrapping, Shapley values, and appropriate distance metrics, respectively, and can be visualized in the same way.

\section{Conclusion}
In conclusion, we have implemented a visual clinical decision support system based on a temporal deep-reinforcement learning model that is capable of simulating patient treatment outcomes. To our knowledge, this is one of the few attempts at an explainable AI focused interface for clinical users, as well as one of the first attempts at a visual interface to explore a dual digital twin system in a healthcare setting. Through our participatory design, we highlight several findings with a focus on balancing information density, usability, and encouraging appropriate trust for a variety of of end users. In our future work, we hope to evaluate this interface on a large range of clinical end users in practice, as well as extend our work to even more detailed decision-making that can consider more patient quality-of-life measures.

\appendix 
\pagebreak
\clearpage
\onecolumn

\noindent \cref{appendixa} contains additional details about the model (\cref{details}) and an evaluation of the deep learning models, along with details about the data cohort (\cref{evaluation}). 
\\
\noindent \cref{appendixb} contains additional figures of the images of related content for the system (\cref{additionalfigures}) and images of earlier prototypes (\cref{prototypes}).

\section{Appendix A: Model Details and Evaluation}\label{appendixa}
\subsection{Model Details}\label{details}

\subsubsection{Patient Simulator Models}
\begin{figure*}[ht]
    \centering
    \includegraphics[width=.9\textwidth, alt={Architecture for the transition and outcome deep survival machines (DSM). Patient state and previous state treatment decision use a standard DNN with input dropout to improve the models ability to deal with unknown data. The decision is concatenated to the penultimate layer in order to prevent the model from relying only on correlated features due to the use of dropout during training. DSM models predict a mixture of model parameters for each patient from a pretrained set of user-defined mixtures.}]{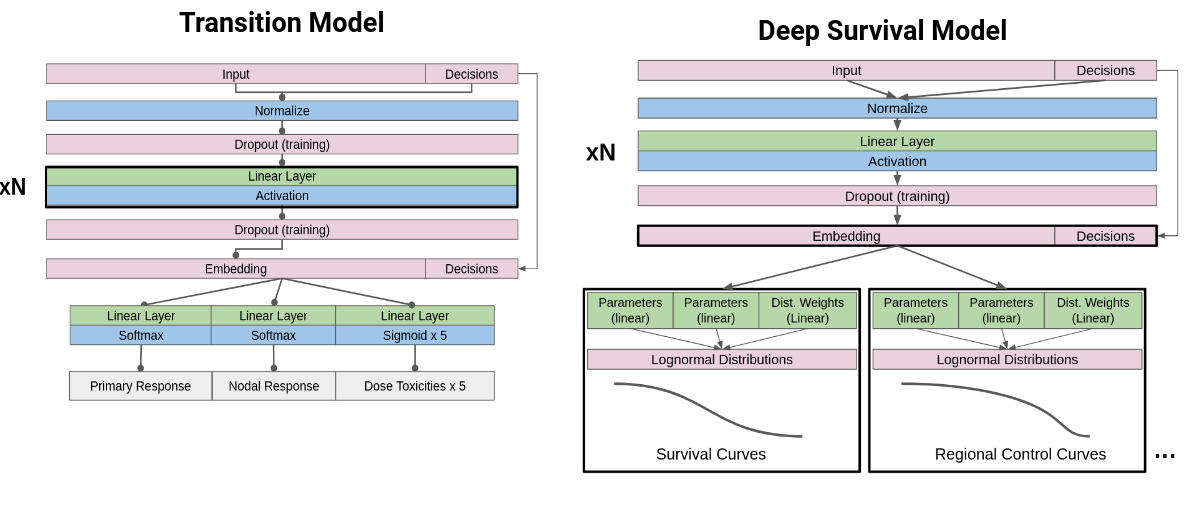}
    \caption{Architecture for the transition and deep survival  models (DSM). Patient state and previous state treatment decision use a standard DNN with input dropout to improve the models ability to deal with unknown data. The decision is concatenated to the penultimate layer in order to prevent the model from relying only on correlated features due to the use of dropout during training. DSM models predict a mixture of model parameters for each patient from a pre-trained set of user-defined number of mixtures.}
    \label{fig:dt_transitionmodel}
\end{figure*}

To simulate the patient, we use a set of models to mimic intermediate response to treatment (transition models), and long-term response after treatment (outcome models). \cref{fig:dt_transitionmodel} Shows the architecture for the transition models and deep survival machines used to model temporal patient outcomes (DSMs). Each time-point uses a separate transition state model. For induction chemotherapy, we constrain the model to not allow for any tumor response when no chemotherapy is given, as this would indicate no treatment at this point.

Transition models predict patient response to treatment in terms of tumor shrinkage and severe toxicities from treatment. Specifically, we consider primary disease response (PD), and nodal disease response (ND), which are each 4 categorical ordinal variables, as well as 5 binary results for different types of dose-limiting toxicities (DLTS). For the case of Induction chemotherapy, disease response is always assumed to be stable when no treatment is done. Seperate models are trained for post-IC and post-CC transitions, as this resulted in better performance.

For the outcome model, two separate models are used. The first is a deep neural network that predicts toxicity risk using binary variables: Aspiration (AS), and Feeding Tube (FT) at 6 months after treatment. 

The second outcome model predicts cumulative patient risk over time for overall survival (OS), locoregional control (LRC), and distant metastases (FDM) for up to 5 years. Temporal risk models use a variant of deep survival machines (DSM)~\cite{nagpal2021deep}. For all three outcomes, the DSM model returns a mixture of parametric log-normal distributions for the patient that can be used to provide a cumulative survival risk over time.

Because clinicians listed confidence intervals as important for reasoning about the model predictions (T3.3), all transition and outcome models are trained using dropout on the penultimate layer between 50\% and 75\%. During evaluation, we re-run each prediction with random dropout at least 20 times, and then save the 95\% confidence intervals for each prediction.~\cite{gal2016dropout}.

All models implemented in pytorch and trained using the Adam optimizer. Models were trained using early stopping until the validation loss stopped increasing for at least 10 epochs. Transition models, static outcome models, and Deep Survival Machines for temporal outcomes used a dropout of 10\% on the input layer and 50\% on the penultimate layer during training. Transition models and static outcome models used 2 hidden layers with an output size of 500 each. The deep survival machines used a single hidden layer with a size of 100 and 6 different distributions for each outcome.

\subsubsection{Policy Models}

\begin{figure}
    \centering
    \includegraphics[width=.6\linewidth, alt={Architecture for policy model used to simulate a physician decision. Both the optimal and imitation models use a shared embedding with a custom position token at each stage, followed by a separate layer for each output, with additional fully connected layers unique to each model before the output. Model activations for the penultimate layers are used when calculating similar patients. Policy models use a modified version of a transformer encoder that saves the cohort at each time point into memory at training time.}]{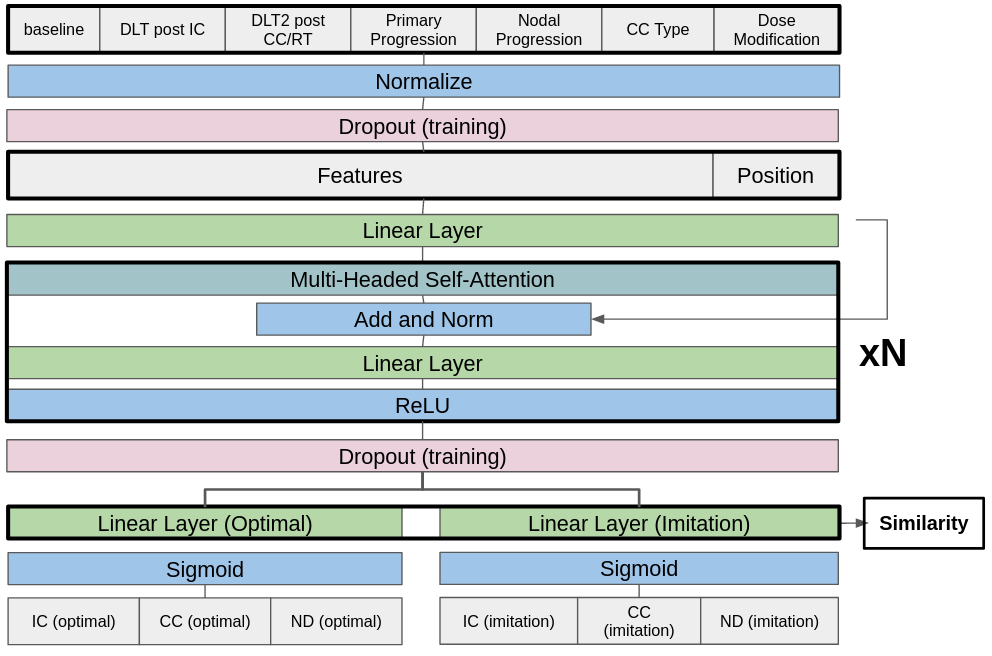}
    \caption{ Architecture for policy model used to simulate a physician decision. Both the optimal and imitation models use a shared embedding with a custom position token at each stage, followed by a separate layer for each output, with additional fully connected layers unique to each model before the output. Model activations for the penultimate layers are used when calculating similar patients. Policy models use a modified version of a transformer encoder that saves the cohort at each time point into memory at training time.}
    \label{fig:dt_models}
\end{figure}

The patient simulator models and ground truth responses are used as the environment to train a digital physician (policy model) (\cref{fig:dt_models}). Because there is disagreement among users as to whether they prefer to see what a physician would do, or what the "best" choice should be, we jointly train two versions of the policy model: one that minimizes a combination of patient risks based on the patient simulator responses (optimal policy model), and one that predicts what a physician would do based on the cohort data (imitation policy model). 

Each policy model (optimal and imitation) is trained using a dual loss function: prediction of the ground truth (or optimal) decision sequence, and triplet loss. Triplet loss is included as it was found to increase model performance in terms of AUC and accuracy for the imitation model. We use AUC as it is a measure of the relative ranking of patient risks, and is thus commonly used to identify rarer events. Specifically, the loss for a given patient $p$ at each epoch for a given output (optimal or imitation) is given by:

$$
L(p) = w_{1}\cdot \sum_{i = 0}^{2}BCE(\hat{y_{p,i}},y_{p,i}) + w_{2}\cdot max\big(d(a_{p},b_{p}) - d(a_{p},c_{p} )+1,0\big)
$$

Where $y_{i}$ and $\hat{y_{i}}$ are the ground truth and predicted decisions, respectively. $d(\cdot,\cdot)$ is the euclidean distance. $a_{p}$ is the final hidden layer weight vector for the patient. $b_{p}$ are the hidden weights for a randomly sampled patient with the same ground truth treatment sequence, and $c_{p}$ is a randomly sampled patient with a different treatment sequence. $w_{1}$ and $w_{2}$ are user-decided weights. For our implementation we use $w_{1} = 1$ and $w_{2} = .2$ for both outputs.

Both optimal and imitation policy models use shared layers until the penultimate layers, which are unique to each output, and are re-trained each epoch (\cref{fig:dt_models}). This allows for joint learning of important features from each other. To encourage our model to explicitly consider other patients in the cohort, our policy model architecture uses a transformer encoder and uses a position token to encode the temporal state of the patient. During evaluation on a new datapoint, the cohort data for the current state is used as the Query input of the multi-headed attention as described in Vaswani et al.~\cite{vaswani2017attention}.

Imitation policy model decisions are trained using the unaltered ground truth states in the data to predict the decision made by the clinician. The optimal model decision is, in contrast, trained on using random data augmentation on the pre-treatment variables for the patient for each epoch. Specifically, each column (feature) has a 25\% probability of being pseudo-randomly shuffled in the training sample, and the predicted patient response to treatment using the deep learning transition models. 

When determining the treatment sequence for the optimal decision, we calculate the decisions that minimize a combination of all predicted outcomes, given by:

$$L = w_{tox}\sum_{z \in Z}w_{z} P(z=1) + w_{s}\sum_{o \in O}\frac{w_{o}}{\tilde{T}(o)}$$

Where $z \in Z$ is the set of binary outcomes (e.g toxicity, 4 year survival, 4 year locoregional control), $o \in O$ is the set of temporal survival outcomes (survival, locoregional control, distant control), $\tilde{T}(o)$ is the median predicted time-to-event of outcome $o$, and $w \in W$ a set of user defined weights for each aspect of the loss function.

Because we need to explain the policy model recommendations (T3), we use integrated gradients~\cite{sundararajan2017axiomatic} to obtain feature importance for each decision relative to a baseline value. Integrated gradients was chosen as it satisfies the completeness axiom where attributions sum to the difference in the prediction between the baseline and actual recommendation, which was found to be easier to reason about with our clinicians. For our baseline, we assume the lowest possible rating for most ordinal attributes such as tumor staging or disease response, and the most common value for categorical attributes such as gender, ethnicity, and tumor subsite, as well as age and dose to the main tumor, based on feedback from clinicians and what they found most intuitive.

All models implemented in pytorch and trained using the Adam optimizer. Models were trained using early stopping until the validation loss stopped increasing for at least 10 epochs. Our policy model used an input dropout of 10\% and 25\% dropout on the final layers, with a single transformer encoder of size 1000 for the joint embedding, and a linear layer of size 20 for both the optimal and imitation model outputs. 

\subsubsection{KNN-based Symptom Prediction}
Our symptom prediction model uses a different cohort of patient and relies on a KNN predictor using the embeddings taken from a model trained to predict symptom trajectories. Specifically, we trained a fully connected deep learning model to predict symptom ratings for each symptom and each time point in the data. Time points considered were at 0, 7, 12, and 27 weeks after starting radiation therapy. Outputs were treated as independent values with a sigmoid loss function that was scaled to be between 0 and 10. Input features were gender, packs-years, HPV status, treatment dose and dose frction, race, tumor laterality, tumor subsite, T-category, N-category, and treatment decisions for induction chemotherapy (IC) and concurrent chemotherapy (CC).

Patient embeddings for the cohort were taken from the model activations in the batch-normalized penultimate layer in the deep learning model. When predicting a new patient, we take the new patient's model embeddings and extract the 10 most similar patients, based on euclidean distance, from the embeddings of both the treated and untreated patients, respectively. Patient symptom profiles are taken for these patient seperately

During deep learning model training we used an 80/20 train validation split on the data for parameter tuning, using the mean-squared-error loss (MSE). Missing symptom values were ignored in the loss function. All models were trained using the ADAM optimizer in pytorch using early stopping on the validation loss. Our final model used a single hidden layer of size 10 with the ReLu activation, followed by batch normalization, with no dropout.

\subsection{Model Evaluation}\label{evaluation}
536 patients were used to evaluate our system. The dataset was split into a training cohort of 389 patients and an evaluation cohort of 147 patients before beginning the development of the models. For evaluation purposes, the training sample was stratified in order to get a minimum of 3 patients with each endpoint, and treatment decision in the model. Because we could not achieve enough samples of patients with several dose limiting toxicities, all toxicities that were not present in both cohorts were aggregated into an "other" category for the purpose of modeling and evaluation. 
The features used for the entire cohort, excluding lymph node patterns, is shown in (\cref{tab:dt_demographics}), stratified by treatment sequence. An anova F-test was used to analyze correlations between each feature set and the treatment sequence, and p-values are included in the table.

Performance of the policy model with and without triplet loss is shown in (\cref{tab:policyextended}). We see an increase in imitation model performance, with a slight decrease in "optimal" model performance for accuracy but increase in AUC. This is likely due to the heavy imbalance in the optimal outcomes: only 10.8\% of cases recommended concurrent chemotherapy and 19\% of cases recommended neck dissection, as rare events were predicted with higher prediction confidence. Given that a majority of users preferred to use the "imitation" model, the triplet model was used in practice. 

In general, AUC tended to perform better than Accuracy in the optimal model, likely due to the heavy imbalance in the optimal outcomes: only 10.8\% of cases recommended concurrent chemotherapy and 19\% of cases recommended neck dissection. In general, model performance is comparable to similar outcome models from earlier studies, considering the added difficulty of optimizing for 23 different outcomes and 6 treatment decisions. Interestingly, our optimal model suggested induction chemotherapy followed by radiation alone a majority of the time, which contradicts the standard practice where concurrent chemotherapy is standard while induction is used for patients with very large tumor spread that needs to be reduced before applying radiation. However, the data is largely limited by confounders and lack of detailed information on how changes in patient's health affect treatment and outcomes. Additionally, we have been told that the specific grade of dose-limiting toxicity is an important factor in treatment and side effects, which our model does not consider.

Performance of transition models are shown in  (\cref{tab:dt_transitionmodels}). Because the outcomes we want to predict are often rare events, we compared default training performance with basic cross-entropy loss with a balanced loss function. Non-balanced models generally performed better in terms of AUC with similar accuracy. 

To evaluate time series models, we calculate F1 and ROC AUC scores at 12, 24, 36, and 48 months after treatment (\cref{tab:dt_dsmpeformance}). We exclude longer periods as we tend to have fewer followup data available after 48 months. OS, FDM and LRC models tend to have high F1 score but modest AUC scores, possibly due to the fact that failures are rare events in the data.

\begin{table}[h]
\centering
\begin{tabular}{|lrrrr|}
\hline
\multicolumn{1}{|l|}{\multirow{2}{*}{Decision}} & \multicolumn{2}{c|}{Optimal}                              & \multicolumn{2}{c|}{Imitation}                            \\ \cline{2-5} 
\multicolumn{1}{|l|}{}                          & \multicolumn{1}{c|}{AUC}  & \multicolumn{1}{c|}{Accuracy} & \multicolumn{1}{c|}{AUC}  & \multicolumn{1}{c|}{Accuracy} \\ \hline

\multicolumn{5}{|c|}{With Triplet Loss}                                                                                                                                 \\ \hline
\multicolumn{1}{|l|}{IC}                        & \multicolumn{1}{r|}{0.84} & \multicolumn{1}{r|}{0.58}     & \multicolumn{1}{r|}{0.79} & 0.88                          \\ \hline
\multicolumn{1}{|l|}{CC}                        & \multicolumn{1}{r|}{0.97} & \multicolumn{1}{r|}{0.73}     & \multicolumn{1}{r|}{0.93} & 0.78                          \\ \hline
\multicolumn{1}{|l|}{ND}                        & \multicolumn{1}{r|}{0.95} & \multicolumn{1}{r|}{0.79}     & \multicolumn{1}{r|}{0.90} & 0.81                          \\ \hline
\multicolumn{5}{|c|}{No Triplet Loss}                                                                                                                                   \\ \hline
\multicolumn{1}{|l|}{IC}                        & \multicolumn{1}{r|}{0.82} & \multicolumn{1}{r|}{0.71}     & \multicolumn{1}{r|}{0.60} & 0.87                          \\ \hline
\multicolumn{1}{|l|}{CC}                        & \multicolumn{1}{r|}{0.96} & \multicolumn{1}{r|}{0.91}     & \multicolumn{1}{r|}{0.74} & 0.78                          \\ \hline
\multicolumn{1}{|l|}{ND}                        & \multicolumn{1}{r|}{0.94} & \multicolumn{1}{r|}{0.88}     & \multicolumn{1}{r|}{0.84} & 0.81                          \\ \hline
\end{tabular}
\label{tab:policyextended}
\caption{Physician Simulator Policy Model Performace with and without use of triplet loss.}
\end{table}

\small
\tabcolsep=0.11cm
    \begin{longtable}{@{}|l|l|l|l|l|l|l|l|l|l|@{}}
    \toprule
    Treatment Sequence            & CC      & None    & CC + ND & IC + CC & IC + CC + ND & IC      & ND      & IC + ND  & P-Value                   \\* \midrule
    \endfirsthead
    \endhead
    Count                         & 223     & 57      & 51      & 100     & 36           & 45      & 11      & 13       & 1                         \\* \midrule
    HPV+                          & 56.50\% & 80.70\% & 54.90\% & 50.00\% & 61.11\%      & 42.22\% & 54.55\% & 61.54\%  & \multirow{2}{*}{6.44E-03} \\* \cmidrule(r){1-9}
    HPV Unknown                   & 6.28\%  & 1.75\%  & 7.84\%  & 16.00\% & 11.11\%      & 2.22\%  & 18.18\% & 7.69\%   &                           \\* \midrule
    Age (Mean)                    & 59.3    & 61.3    & 57.7    & 58.5    & 58.3         & 57.6    & 59.6    & 57.0     & 4.92E-01                  \\* \midrule
    Pack-years                    & 17.6    & 10.5    & 18.9    & 17.6    & 21.8         & 15.4    & 16.7    & 4.8      & 1.83E-01                  \\* \midrule
    Male                          & 88.34\% & 80.70\% & 92.16\% & 87.00\% & 91.67\%      & 88.89\% & 81.82\% & 92.31\%  & 6.76E-01                  \\* \midrule
    Smoker                        & 19.28\% & 19.30\% & 35.29\% & 22.00\% & 22.22\%      & 24.44\% & 18.18\% & 0.00\%   & \multirow{2}{*}{2.38E-01} \\* \cmidrule(r){1-9}
    Former Smoker                 & 42.15\% & 40.35\% & 29.41\% & 34.00\% & 33.33\%      & 33.33\% & 54.55\% & 30.77\%  &                           \\* \midrule
    Bilateral                     & 4.48\%  & 3.51\%  & 5.88\%  & 4.00\%  & 2.78\%       & 2.22\%  & 0.00\%  & 0.00\%   & 9.50E-01                  \\* \midrule
    T-category\_1                 & 18.83\% & 63.16\% & 5.88\%  & 6.00\%  & 13.89\%      & 28.89\% & 54.55\% & 30.77\%  & 1.30E-18                  \\* \midrule
    T-category\_2                 & 42.15\% & 33.33\% & 54.90\% & 33.00\% & 27.78\%      & 48.89\% & 45.45\% & 61.54\%  & 4.45E-02                  \\* \midrule
    T-category\_3                 & 24.66\% & 3.51\%  & 21.57\% & 29.00\% & 27.78\%      & 17.78\% & 0.00\%  & 7.69\%   & 3.25E-03                  \\* \midrule
    T-category\_4                 & 14.35\% & 0.00\%  & 17.65\% & 32.00\% & 30.56\%      & 4.44\%  & 0.00\%  & 0.00\%   & 6.69E-08                  \\* \midrule
    N-category\_1                 & 52.91\% & 80.70\% & 52.94\% & 27.00\% & 16.67\%      & 22.22\% & 63.64\% & 61.54\%  & 4.96E-14                  \\* \midrule
    N-category\_2                 & 39.46\% & 12.28\% & 43.14\% & 65.00\% & 75.00\%      & 73.33\% & 27.27\% & 38.46\%  & 7.13E-14                  \\* \midrule
    N-category\_3                 & 1.79\%  & 0.00\%  & 0.00\%  & 8.00\%  & 8.33\%       & 4.44\%  & 0.00\%  & 0.00\%   & 1.91E-02                  \\* \midrule
    AJCC\_2                       & 15.25\% & 5.26\%  & 15.69\% & 16.00\% & 22.22\%      & 22.22\% & 9.09\%  & 7.69\%   & 1.66E-16                  \\* \midrule
    AJCC\_3                       & 9.42\%  & 5.26\%  & 13.73\% & 22.00\% & 25.00\%      & 0.00\%  & 18.18\% & 0.00\%   & 2.10E-04                  \\* \midrule
    AJCC\_4                       & 36.77\% & 12.28\% & 39.22\% & 49.00\% & 38.89\%      & 57.78\% & 18.18\% & 38.46\%  & 5.90E-05                  \\* \midrule
    subsite\_BOT                  & 50.22\% & 35.09\% & 47.06\% & 56.00\% & 55.56\%      & 57.78\% & 18.18\% & 46.15\%  & 7.85E-02                  \\* \midrule
    subsite\_GPS                  & 0.90\%  & 1.75\%  & 1.96\%  & 2.00\%  & 8.33\%       & 0.00\%  & 0.00\%  & 7.69\%   & 7.50E-02                  \\* \midrule
    subsite\_Soft palate          & 0.90\%  & 1.75\%  & 3.92\%  & 1.00\%  & 0.00\%       & 0.00\%  & 0.00\%  & 0.00\%   & 6.48E-01                  \\* \midrule
    subsite\_Tonsil               & 41.26\% & 54.39\% & 41.18\% & 36.00\% & 33.33\%      & 40.00\% & 81.82\% & 30.77\%  & 4.78E-02                  \\* \midrule
    Pathological Grade\_1         & 0.90\%  & 0.00\%  & 0.00\%  & 3.00\%  & 0.00\%       & 0.00\%  & 9.09\%  & 0.00\%   & 1.04E-01                  \\* \midrule
    Pathological Grade\_2         & 28.25\% & 31.58\% & 27.45\% & 28.00\% & 33.33\%      & 28.89\% & 45.45\% & 7.69\%   & 6.63E-01                  \\* \midrule
    Pathological Grade\_3         & 50.67\% & 54.39\% & 56.86\% & 48.00\% & 55.56\%      & 46.67\% & 36.36\% & 61.54\%  & 8.39E-01                  \\* \midrule
    Pathological Grade\_4         & 0.90\%  & 0.00\%  & 0.00\%  & 0.00\%  & 0.00\%       & 0.00\%  & 0.00\%  & 7.69\%   & 5.00E-02                  \\* \midrule
    White/Caucasion               & 93.27\% & 89.47\% & 96.08\% & 86.00\% & 86.11\%      & 93.33\% & 90.91\% & 92.31\%  & 3.57E-01                  \\* \midrule
    Aspiration Pre-Therapy        & 2.24\%  & 0.00\%  & 1.96\%  & 7.00\%  & 5.56\%       & 2.22\%  & 0.00\%  & 0.00\%   & 2.14E-01                  \\* \midrule
    Total Dose (gy)               & 68.99   & 66.86   & 69.47   & 69.36   & 69.33        & 67.42   & 68.05   & 67.23    & 1.05E-14                  \\* \midrule
    Dose Fractions                & 2.10    & 2.16    & 2.08    & 2.11    & 2.08         & 2.15    & 2.17    & 2.18     & 7.82E-05                  \\* \midrule
    Survival (Months)             & 76.26   & 71.24   & 80.52   & 74.10   & 74.47        & 87.80   & 98.57   & 97.21    & 1.76E-02                  \\* \midrule
    Locoregional control (Months) & 74.46   & 67.63   & 67.28   & 71.36   & 63.85        & 86.54   & 72.10   & 94.07    & 2.50E-02                  \\* \midrule
    FDM (months)                  & 74.69   & 71.00   & 77.95   & 72.38   & 69.56        & 84.34   & 94.86   & 97.21    & 5.02E-02                  \\* \midrule
    Overall Survival              & 75.34\% & 82.46\% & 70.59\% & 75.00\% & 61.11\%      & 86.67\% & 63.64\% & 100.00\% & 4.36E-02                  \\* \midrule
    Locoregional Control          & 91.03\% & 92.98\% & 68.63\% & 85.00\% & 69.44\%      & 91.11\% & 63.64\% & 84.62\%  & 1.74E-05                  \\* \midrule
    FDM                           & 89.69\% & 96.49\% & 86.27\% & 90.00\% & 80.56\%      & 88.89\% & 72.73\% & 100.00\% & 1.25E-01                  \\* \midrule
    FT                            & 17.49\% & 5.26\%  & 21.57\% & 25.00\% & 38.89\%      & 8.89\%  & 18.18\% & 0.00\%   & 4.80E-04                  \\* \midrule
    Aspiration Post-Therapy       & 17.49\% & 3.51\%  & 25.49\% & 22.00\% & 41.67\%      & 8.89\%  & 18.18\% & 7.69\%   & 1.87E-04                  \\* \midrule
    CR Primary                    & 0.00\%  & 0.00\%  & 0.00\%  & 34.00\% & 38.89\%      & 66.67\% & 0.00\%  & 46.15\%  & 2.99E-50                  \\* \midrule
    PR Primary                    & 0.00\%  & 0.00\%  & 0.00\%  & 52.00\% & 55.56\%      & 28.89\% & 0.00\%  & 30.77\%  & 2.02E-51                  \\* \midrule
    CR Nodal                      & 0.00\%  & 0.00\%  & 0.00\%  & 10.00\% & 2.78\%       & 11.11\% & 0.00\%  & 0.00\%   & 1.79E-06                  \\* \midrule
    PR Nodal                      & 0.00\%  & 0.00\%  & 0.00\%  & 75.00\% & 88.89\%      & 86.67\% & 0.00\%  & 84.62\%  & 1.21E-148                 \\* \midrule
    DLT after CC                  & 0.00\%  & 0.00\%  & 0.00\%  & 75.00\% & 50.00\%      & 64.44\% & 0.00\%  & 84.62\%  & 0.00E+00                  \\* \midrule
    CR Primary 2                  & 83.41\% & 91.23\% & 68.63\% & 90.00\% & 72.22\%      & 91.11\% & 81.82\% & 92.31\%  & 4.86E-03                  \\* \midrule
    PR Primary 2                  & 16.14\% & 7.02\%  & 25.49\% & 10.00\% & 19.44\%      & 4.44\%  & 18.18\% & 7.69\%   & 3.56E-02                  \\* \midrule
    CR Nodal 2                    & 52.02\% & 52.63\% & 17.65\% & 58.00\% & 19.44\%      & 57.78\% & 9.09\%  & 7.69\%   & 1.38E-09                  \\* \midrule
    PR Nodal 2                    & 43.95\% & 35.09\% & 80.39\% & 34.00\% & 77.78\%      & 37.78\% & 90.91\% & 69.23\%  & 4.46E-11                  \\* \midrule
    DLT After CC/RT               & 27.80\% & 0.00\%  & 15.69\% & 29.00\% & 25.00\%      & 0.00\%  & 0.00\%  & 0.00\%   & 2.95E-07                  \\* \bottomrule
    
    \label{tab:dt_demographics}
    \end{longtable}

\small
    \begin{table}[]
    \centering
    \begin{tabular}{@{}llll@{}}
    \toprule
    State                                 & Outcome                 & Metric        & Value \\ \midrule
    \multirow{17}{*}{After IC}            & Primary Response        & accuracy      & 0.404 \\ \cmidrule(l){2-4} 
                                          & Primary Response        & auc\_micro    & 0.801 \\ \cmidrule(l){2-4} 
                                          & Primary Response        & auc\_weighted & 0.674 \\ \cmidrule(l){2-4} 
                                          & Nodal Response          & accuracy      & 0.333 \\ \cmidrule(l){2-4} 
                                          & Nodal Response          & auc\_micro    & 0.853 \\ \cmidrule(l){2-4} 
                                          & Nodal Response          & auc\_weighted & 0.533 \\ \cmidrule(l){2-4} 
                                          & Dose Modification       & accuracy      & 0.333 \\ \cmidrule(l){2-4} 
                                          & DLT\_Gastrointestinal   & accuracy      & 0.804 \\ \cmidrule(l){2-4} 
                                          & DLT\_Other              & accuracy      & 0.946 \\ \cmidrule(l){2-4} 
                                          & DLT\_Dermatological     & accuracy      & 0.893 \\ \cmidrule(l){2-4} 
                                          & DLT\_Hematological      & accuracy      & 0.786 \\ \cmidrule(l){2-4} 
                                          & DLT\_Neurological       & accuracy      & 0.911 \\ \cmidrule(l){2-4} 
                                          & DLT\_Gastrointestinal   & auc           & 0.497 \\ \cmidrule(l){2-4} 
                                          & DLT\_Other              & auc           & 0.415 \\ \cmidrule(l){2-4} 
                                          & DLT\_Dermatological     & auc           & 0.420 \\ \cmidrule(l){2-4} 
                                          & DLT\_Hematological      & auc           & 0.511 \\ \cmidrule(l){2-4} 
                                          & DLT\_Neurological       & auc           & 0.557 \\ \midrule
    \multirow{16}{*}{After RT + CC}       & Primary Response        & accuracy      & 0.333 \\ \cmidrule(l){2-4} 
                                          & Primary Response        & auc\_micro    & 0.887 \\ \cmidrule(l){2-4} 
                                          & Primary Response        & auc\_weighted & 0.568 \\ \cmidrule(l){2-4} 
                                          & Nodal Response          & accuracy      & 0.372 \\ \cmidrule(l){2-4} 
                                          & Nodal Response          & auc\_micro    & 0.756 \\ \cmidrule(l){2-4} 
                                          & Nodal Response          & auc\_weighted & 0.545 \\ \cmidrule(l){2-4} 
                                          & DLT\_Gastrointestinal   & accuracy      & 0.918 \\ \cmidrule(l){2-4} 
                                          & DLT\_Other              & accuracy      & 0.980 \\ \cmidrule(l){2-4} 
                                          & DLT\_Dermatological     & accuracy      & 0.966 \\ \cmidrule(l){2-4} 
                                          & DLT\_Hematological      & accuracy      & 0.952 \\ \cmidrule(l){2-4} 
                                          & DLT\_Neurological       & accuracy      & 0.966 \\ \cmidrule(l){2-4} 
                                          & DLT\_Gastrointestinal   & auc           & 0.564 \\ \cmidrule(l){2-4} 
                                          & DLT\_Other              & auc           & 0.727 \\ \cmidrule(l){2-4} 
                                          & DLT\_Dermatological     & auc           & 0.625 \\ \cmidrule(l){2-4} 
                                          & DLT\_Hematological      & auc           & 0.613 \\ \cmidrule(l){2-4} 
                                          & DLT\_Neurological       & auc           & 0.552 \\ \midrule
    \multirow{10}{*}{After All Treatment} & Feeding Tube            & accuracy      & 0.803 \\ \cmidrule(l){2-4} 
                                          & Feeding Tube            & auc           & 0.683 \\ \cmidrule(l){2-4} 
                                          & Feeding Tube            & f1            & 0.216 \\ \cmidrule(l){2-4} 
                                          & Aspiration Post-therapy & accuracy      & 0.803 \\ \cmidrule(l){2-4} 
                                          & Aspiration Post-therapy & auc           & 0.775 \\ \cmidrule(l){2-4} 
                                          & Aspiration Post-therapy & f1            & 0.065 \\ \bottomrule
    \end{tabular}
    \label{tab:dt_transitionmodels}
    \caption{Model Performance for all transition states and toxicity}
    \end{table}

\small
    \begin{table}[h]
    \centering
    \begin{tabular}{@{}llll@{}}
    \toprule
    outcome                               & months              & metric & value \\ \midrule
    \multirow{8}{*}{OS}                   & \multirow{2}{*}{12} & AUC    & 0.52  \\ \cmidrule(l){3-4} 
                                          &                     & F1     & 0.99  \\ \cmidrule(l){2-4} 
                                          & \multirow{2}{*}{24} & AUC    & 0.63  \\ \cmidrule(l){3-4} 
                                          &                     & F1     & 0.96  \\ \cmidrule(l){2-4} 
                                          & \multirow{2}{*}{36} & AUC    & 0.64  \\ \cmidrule(l){3-4} 
                                          &                     & F1     & 0.95  \\ \cmidrule(l){2-4} 
                                          & \multirow{2}{*}{48} & AUC    & 0.60  \\ \cmidrule(l){3-4} 
                                          &                     & F1     & 0.94  \\ \midrule
    \multirow{8}{*}{Locoregional Control} & \multirow{2}{*}{12} & AUC    & 0.64  \\ \cmidrule(l){3-4} 
                                          &                     & F1     & 0.97  \\ \cmidrule(l){2-4} 
                                          & \multirow{2}{*}{24} & AUC    & 0.56  \\ \cmidrule(l){3-4} 
                                          &                     & F1     & 0.94  \\ \cmidrule(l){2-4} 
                                          & \multirow{2}{*}{36} & AUC    & 0.57  \\ \cmidrule(l){3-4} 
                                          &                     & F1     & 0.93  \\ \cmidrule(l){2-4} 
                                          & \multirow{2}{*}{48} & AUC    & 0.57  \\ \cmidrule(l){3-4} 
                                          &                     & F1     & 0.92  \\ \midrule
    \multirow{8}{*}{Distant Control}      & \multirow{2}{*}{12} & AUC    & 0.42  \\ \cmidrule(l){3-4} 
                                          &                     & F1     & 0.98  \\ \cmidrule(l){2-4} 
                                          & \multirow{2}{*}{24} & AUC    & 0.62  \\ \cmidrule(l){3-4} 
                                          &                     & F1     & 0.95  \\ \cmidrule(l){2-4} 
                                          & \multirow{2}{*}{36} & AUC    & 0.62  \\ \cmidrule(l){3-4} 
                                          &                     & F1     & 0.93  \\ \cmidrule(l){2-4} 
                                          & \multirow{2}{*}{48} & AUC    & 0.57  \\ \cmidrule(l){3-4} 
                                          &                     & F1     & 0.92  \\ \bottomrule
    \end{tabular}
    \caption{Model Performance for deep survival machines at 12, 24, 36, and 48 months.}
    \label{tab:dt_dsmpeformance}
    \end{table}

\normalsize
\begin{figure*}[ht]
    \centering
    \includegraphics[width=\textwidth, alt={Plots of performance metrics for all models in the system. See tables for actual numbers}]{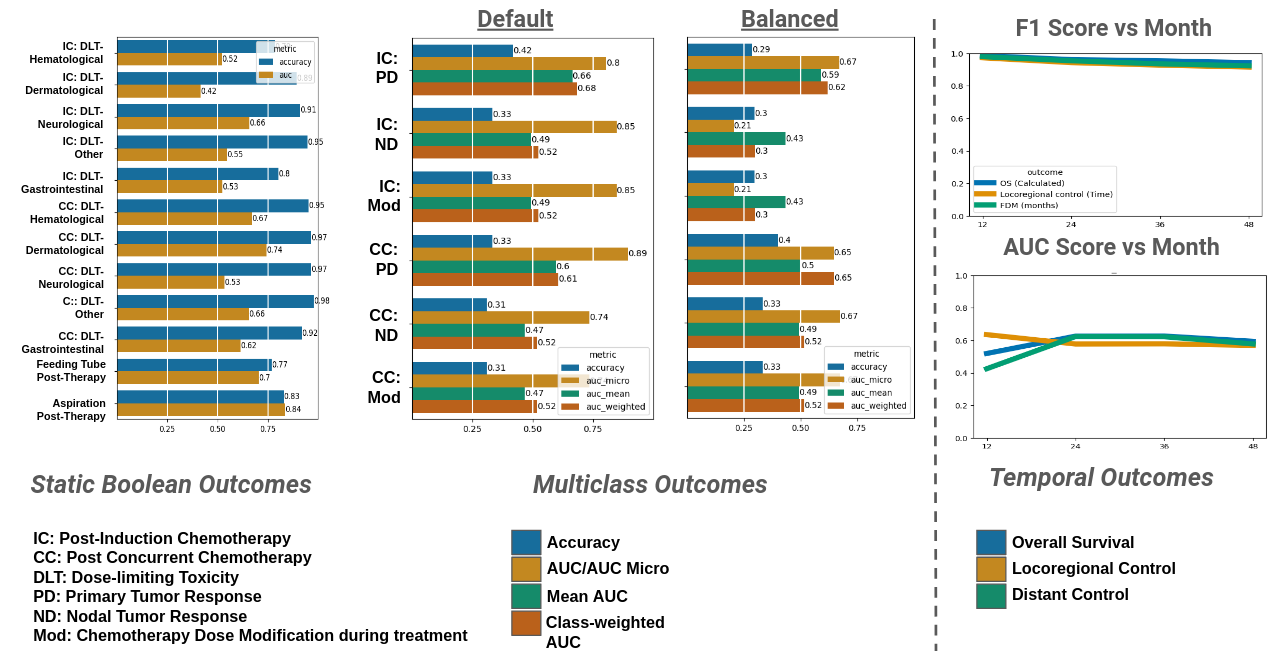}
    \caption{All Transition State Outcomes. (Right) Accuracy and AUC score for boolean outcomes such as toxicities. Models perform well in terms of accuracy and late toxicity (FT and Aspiration), but have mixed AUC results for dose-limiting toxicities due to the heavy imbalance in the data and low number of positive samples to learn from. (Center) Model performance for multi-class transition states (disease response and dose modification) using accuracy and micro, macro, and weighted AUC score for both unweighted and balanced loss weights. Models perforce best in terms of macro AUC score. Balanced models generally performed worse. (Right) F1 score and AUC score for temporal outcomes at 12, 24, 36, and 48 months after treatment. F1 scores tend to be very high while AUC scores stay around .6, likely due to issue with imbalanced data and incomplete censoring.}
    \label{fig:dt_transitionoutcomes}
\end{figure*}

\clearpage
\newpage
\section{Appendix B: Prototypes and Additional Figures}\label{appendixb}
\cref{additionalfigures} Shows additional figures from the interface that were removed for space. \cref{prototypes} Shows earlier designs of the interface. \cref{fig:dt_workflow} Shows a diagram of the user workflow the system was design around.

\subsection{Additional Figures}\label{additionalfigures}
\begin{figure*}[!htb]
    \centering
    \includegraphics[width=.6\textwidth, alt={Scatterplot used during model development using the 2 principle components of the policy model embeddings of the cohort and main patient. Outer and inner color encodes model prediction and ground truth treatment, respectively. Hue is used to differentiate the current patient, the most similar patients, and the rest of the cohort.}]{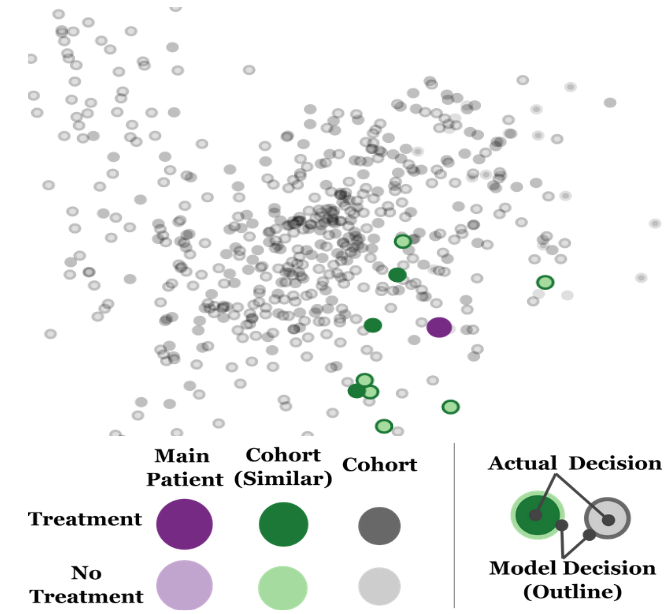}
    \caption{Scatterplot used during model development using the 2 principal components of the policy model embeddings of the cohort and main patient. Outer and inner color encodes model predicted treatment and ground truth treatment, respectively. Hue is used to differentiate the current patient, the most similar patients, and the rest of the cohort.}
    \label{fig:dt_scatterplot}
\end{figure*}

\begin{figure*}[!htb]
    \centering
    \includegraphics[width=.8\textwidth, alt={Simple anatomical diagrams used for spatial features in the visualization. (Left) Dose-limiting toxicities, showing a simple representation of a person with different parts of their body, such as blood vessels, lungs, heart, brain, kidneys, and stomach. (Center) Lymph node regional involvement, with 8 regions on each side of the head. (Right) Primary tumor subsites, includings base of tongue, glossopharyngeal sulcus, pharnygeal wall, and soft palate, with a general outline of the side of the head around the region for reference. All regions not included in the diagram are considered "Not Otherwise Specified".}]{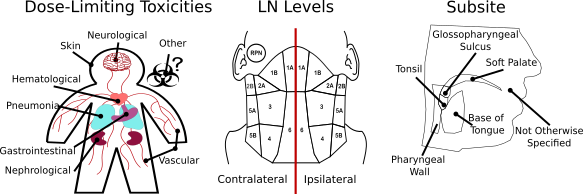}
    \caption{Diagrams used for spatial features in the visualization. (Left) Dose-limiting toxicities. (Center) Lymph node regional involvement. (Right) Primary tumor subsite. All regions not included in the diagram are considered "Not Otherwise Specified".}
    \label{fig:dt_diagrams}
\end{figure*}

\begin{figure*}[!htb]
    \centering
    \includegraphics[width=.8\textwidth, alt={Full figure showing user input buttons, which consists of toggles for each feature, with model toggles at the top. Most buttons are white with a faint outline. buttons of current features are in grey with a faint outline, and buttons for new inputs that haven't yet been sent to the model are white with a large border. At the bottom are diagrams of the head and neck to annotate tumor location and tumor subsites. Diagrams are color coded based on importance to the model treatment recommendation.}]{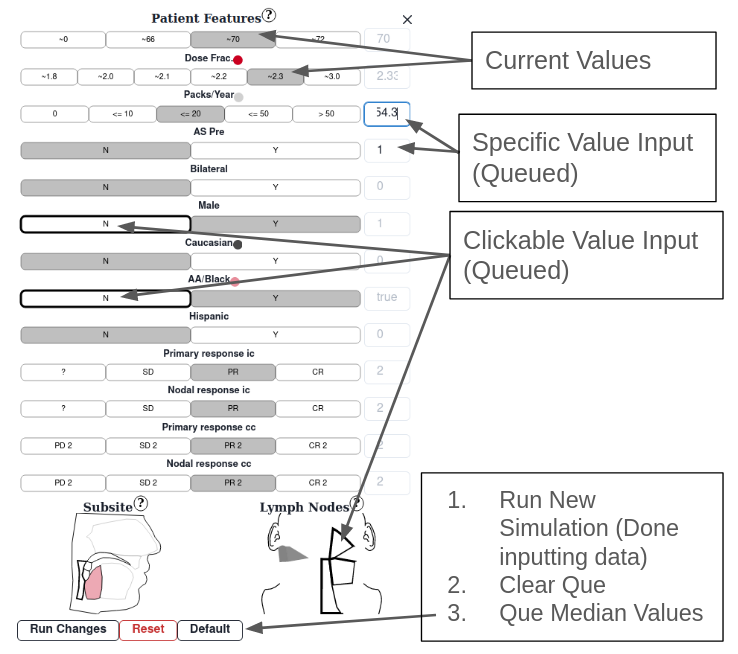}
    \caption{Full view of the user input panel}
    \label{fig:dt_fullinput}
\end{figure*}

\begin{figure*}[!htb]
    \centering
    \includegraphics[width=.9\textwidth, alt={Diagram of user workflow using the interface. (Left) Users start by inputting patient features and setting the model parameters, including the treatment to be considered. (Center) Users start by viewing the most prominent information: treatment recommendations and long-term patient outcome risk plots for survival and disease recurrence. (Right) Users who wish for more information can view additional views such as model explanations, similar patients, and additional patient risk prediction results.}]{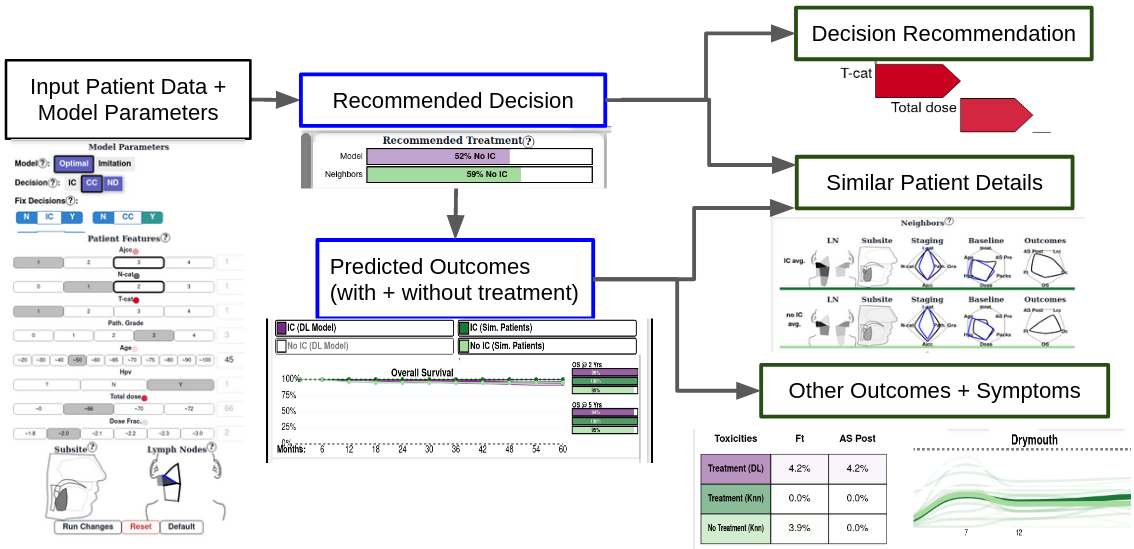}
    \caption{Diagram of user workflow using the interface. (Left) Users start by inputting patient features and setting the model parameters, including the treatment to be considered. (Center) Users start by viewing the most prominent information: treatment recommendations and long-term patient outcome risk plots for survival and disease recurrence. (Right) Users who wish for more information can view additional views such as model explanations, similar patients, and additional patient risk prediction results.}
    \label{fig:dt_workflow}
\end{figure*}

\begin{figure*}[!htb]
    \centering
    \includegraphics[width=.8\textwidth, alt={Pseudocode for the method of estimating average treatment effect for a patient. See main text for description of the algorithm}]{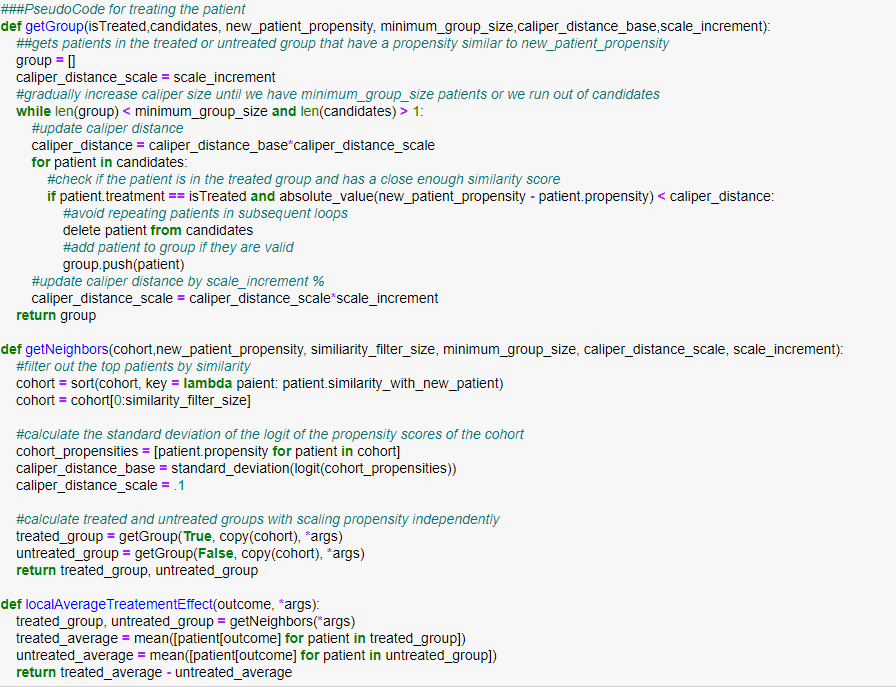}
    \caption{Pseudocode for the method of estimating average treatment effect for a patient}
    \label{fig:dt_atecode}
\end{figure*}

\cref{fig:dt_scatterplot} Shows a scatterplot in the interface of the patient policy model embeddings, which was used during model development to explore the predictions in the cohort and find example patients to test the interface on. The scatterplot is not in the main manuscript as it does not contribute to the clinical user goals. 
\cref{fig:dt_diagrams} shows the spatial feature diagrams for Dose-limiting toxicities, lymph node levels, and tumor subsites used in the interface. 
\cref{fig:dt_fullinput} shows the full patient input panel.
\cref{fig:dt_workflow} Shows a diagram of the user workflow for the system.
\cref{fig:dt_atecode} Shows the psuedocode for the ATE patient sampling algorithm. 

\newpage
\subsection{Prototypes}\label{prototypes}
\begin{figure*}[!htb]
    \centering
    \includegraphics[width=\textwidth, alt={Screenshot of earliest version of the interface. In this version we used a different encoding for treatment recommendation that users found intuitive as it showed the raw model output as percentage of confidence in the patient receiving treatment. This version also showed an additional histogram of the mahalanobis distances for the cohort. We also used a different colorscheme. Additionally, outcomes where shown only as barcharts with a toggle to change the set of outcomes being shown (transition states, dlts, or 4 year post-treatment outcomes). Model parameters where shown at the top instead of alongside the patient panel.}]{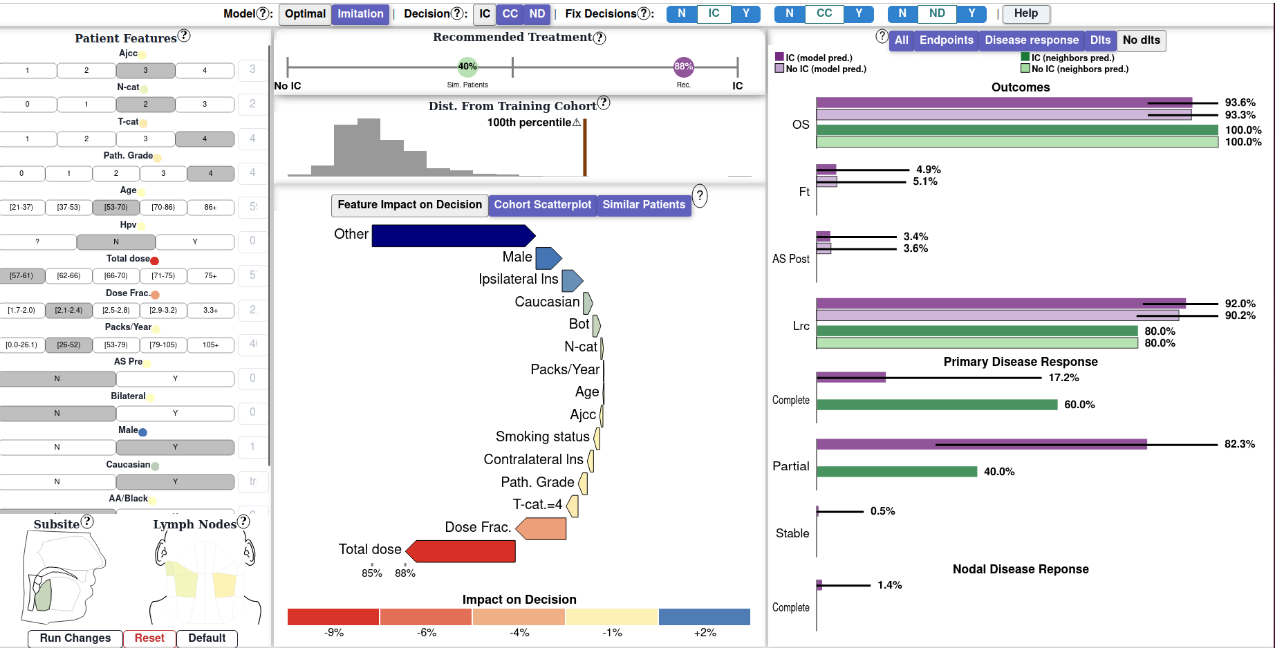}
    \caption{Early version of the interface before integrating temporal outcomes. In this version we used a different encoding for treatment recommendation that users found intuitive as it showed the raw model output as percentage of confidence in the patient receiving treatment. This version also showed an additional histogram of the mahalanobis distances for the cohort. We also used a different colorscheme. Additionally, outcomes where shown only as barcharts with a toggle to change the set of outcomes being shown (transition states, dlts, or 4 year post-treatment outcomes). Model parameters where shown at the top instead of alongside the patient panel.}
    \label{fig:dt_prototype1}
\end{figure*}

\begin{figure*}[!htb]
    \centering
    \includegraphics[width=\textwidth, alt={Screenshot of early version of the interface. In this version the patient input panel was hidden in a "drawer" and could be pulled out via the grey section on the far left, once an initial patient was input. This version includes barcharts with alternative patient outcomes alongside temporal outcomes. Model parameters where shown at the top instead of alongside the patient panel.}]{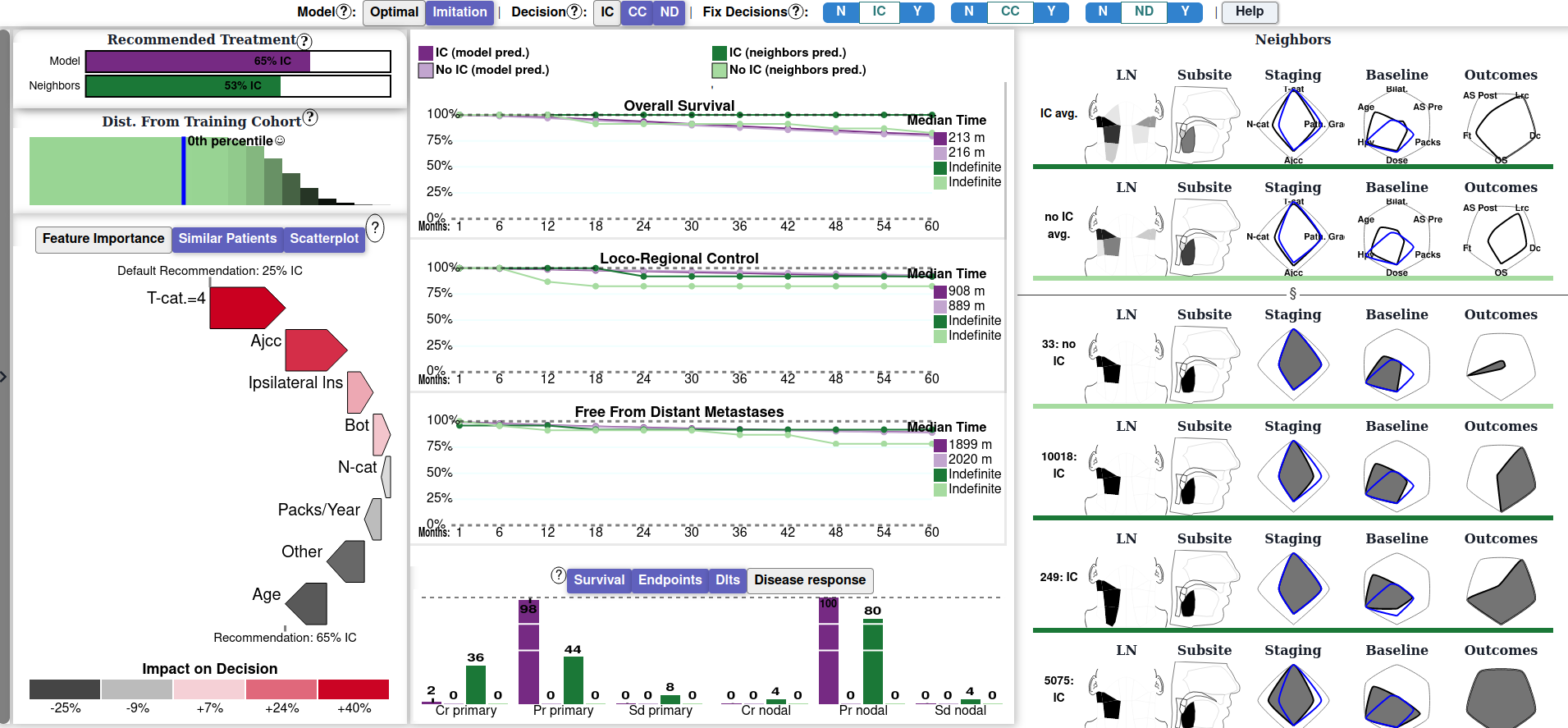}
    \caption{Early version of the interface before the workshop. In this version the patient input panel was hidden in a "drawer" and could be pulled out via the grey section on the far left, once an initial patient was input. This version includes barcharts with alternative patient outcomes alongside temporal outcomes. Model parameters where shown at the top instead of alongside the patient panel.}
    \label{fig:dt_prototype2}
\end{figure*}

\begin{figure*}[!htb]
    \centering
    \includegraphics[width=\textwidth, alt={Screenshot of early version of the outcomes view, which relied on barchars}]{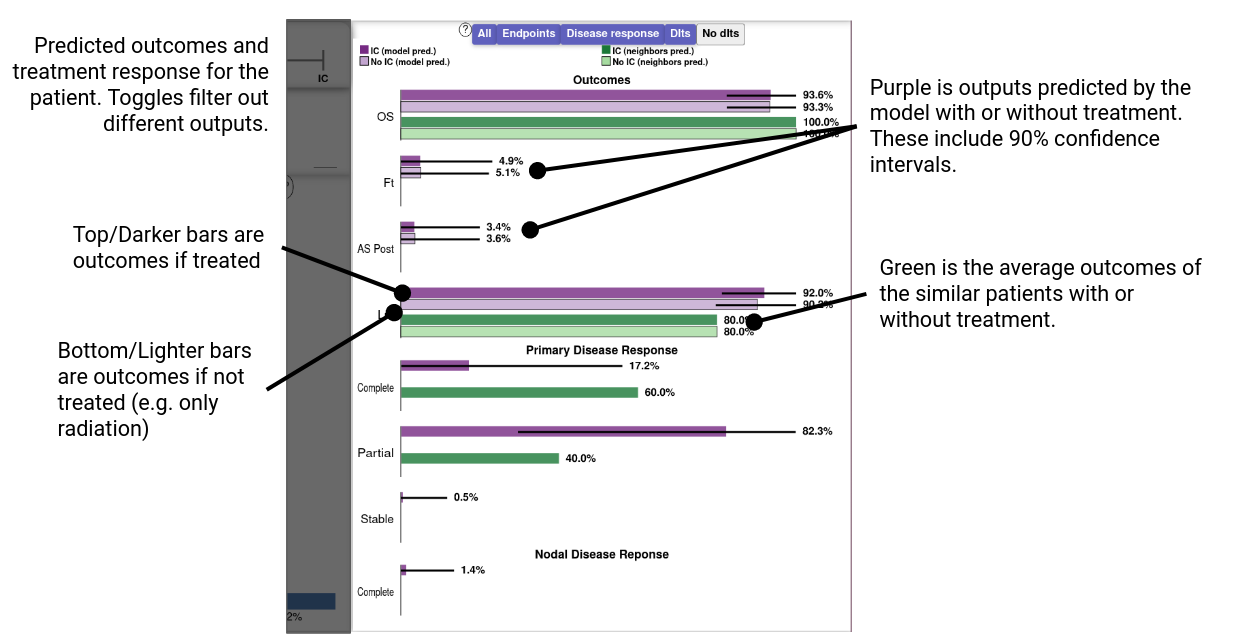}
    \caption{Early version of the outcome view. Our original variant used only static outcomes (4 year survival etc) and focused on barcharts of multiple symptoms, based on the original DT model which used binary outcomes only. This was altered after clinicians states that they were used to dealing with temporal risk plots when reasoning about risk profiles, which also required the addition of the Deep survival machine outcome models.}
    \label{fig:dt_outcome_prototype}
\end{figure*}

\cref{fig:dt_prototype1} and \cref{fig:dt_prototype2} show early versions of the interface. \cref{fig:dt_outcome_prototype} Shows an early version of the outcomes view in more detail. 

\newpage
\clearpage
\twocolumn
\bibliographystyle{abbrv-doi-hyperref-narrow}

\bibliography{template}
\end{document}